\documentclass[10pt]{article}
\usepackage{amsfonts}
\usepackage{amsmath}
\usepackage{amssymb}
\usepackage{mathptmx}
\usepackage{geometry}
\usepackage{graphicx}
\usepackage{hyperref}
\usepackage{cancel}
\usepackage{textcomp}
\usepackage{bm}
\usepackage{times}
\usepackage{epsfig}
\usepackage{xcolor}
\usepackage{slashed}
\usepackage{booktabs}
\usepackage{latexsym}
\usepackage{verbatim}
\usepackage{extarrows}
\usepackage{multirow}
\usepackage{ulem}

\newcommand{\Br}{{\rm Br}}
\newcommand{\capt}{\rm capt}

\newcommand{\GeV}{{\rm GeV}}
\newcommand{\T}{\rm T}
\newcommand{\TeV}{{\rm TeV}}
\newcommand{\MeV}{{\rm MeV}}
\newcommand{\NH}{\rm NH}
\newcommand{\IH}{\rm IH}

\begin{document}
\baselineskip=16pt

\pagenumbering{arabic}

\vspace{1.0cm}

\begin{center}
{\Large\sf Charged lepton flavor-violating transitions in color octet model}
\\[10pt]
\vspace{.5 cm}

{Bin Li~$^{a}$\footnote{libinae@mail.nankai.edu.cn}, Yi Liao~$^{a,b,c}$\footnote{liaoy@nankai.edu.cn} and Xiao-Dong Ma~$^{a}$\footnote{maxid@mail.nankai.edu.cn}}

{
$^a$~School of Physics, Nankai University, Tianjin 300071, China
\\
$^b$ CAS Key Laboratory of Theoretical Physics, Institute of Theoretical Physics,
Chinese Academy of Sciences, Beijing 100190, China
\\
$^c$ Center for High Energy Physics, Peking University, Beijing 100871, China}

\vspace{2.0ex}

{\bf Abstract}

\end{center}

We study charged lepton flavor-violating (LFV) transitions in the color octet model that generates neutrino mass and lepton mixing at one loop. By taking into account neutrino oscillation data and assuming octet particles of TeV scale mass, we examine the feasibility to detect these transitions in current and future experiments. We find that for general values of parameters the branching ratios for LFV decays of the Higgs and $Z$ bosons are far below current and even future experimental bounds. For LFV transitions of the muon, the present bounds can be satisfied generally, while future sensitivities could distinguish between the singlet and triplet color-octet fermions. The triplet case could be ruled out by future $\mu-e$ conversion in nuclei, and for the singlet case the conversion and the decays $\mu\to 3e,~e\gamma$ play complementary roles in excluding relatively low mass regions of the octet particles.

\newpage

\section{Introduction}

Although neutrino oscillations indicate that neutrinos are massive and can change their flavor in weak interactions, no flavor-violating transitions have been observed in the sector of charged leptons. Since the Standard Model (SM) that minimally incorporates neutrino mass and mixing allows those transitions at an extremely small level, the experimental observation of any such type of processes will be a clear imprint of physics beyond SM. These lepton flavor-violating (LFV) processes can be classified into high energy ones that are detected at colliders, such as the LFV decays of the Higgs $h$ and $Z$ bosons, and low energy ones such as $\mu-e$ conversion in nuclei, rare radiative and pure leptonic decays of the $\mu$ and $\tau$ leptons. There are already stringent experimental constraints on some low energy processes: $\mathrm{Br}(\mu\to e\gamma)<4.2\times 10^{-13}$ from MEG \cite{TheMEG:2016wtm}, $\mathrm{Br}(\mu\to 3e)<1.0\times 10^{-12}$ from SINDRUM \cite{Bellgardt:1987du}, and $\mathrm{Br}(\mu \text{Ti}\to e\text{Ti})<4.3\times 10^{-12}$ from SINDRUM II \cite{Dohmen:1993mp}. Significant improvements are expected in the future for some of the processes. The MEG Collaboration has announced plans to reach a sensitivity in the branching ratio as low as $6\times 10^{-14} \cite{Baldini:2013ke}$, while improvements are also anticipated for the $\tau$ lepton decays from searches in $B$ factories~\cite{Aushev:2010bq,Bevan:2014iga}. There are several proposals concerning $\mu-e$ conversion in nuclei whose sensitivities are expected to reach a level ranging from $10^{-14}$ to $10^{-18}$~\cite{Barlow:2011zza,Litchfield:2014qea,H. Natori}. Compared with these low-energy processes, the experimental limits set by colliders are relatively weak, for instance, $\mathrm{Br}(h\to\mu\tau)<0.84\times 10^{-2}$ from CMS \cite{Khachatryan:2015kon}, $\mathrm{Br}(Z\to e\tau)<9.8\times 10^{-6}$ and $\mathrm{Br}(Z\to \mu\tau)<1.2\times 10^{-5}$ from LEP \cite{Abreu:1996mj}. For reference, we collect in Table \ref{tab:experimental_data} the present experimental bounds and expected sensitivities for the above LFV processes involving charged leptons.

Any new particles and interactions that generate neutrino mass and mixing generically induce LFV transitions in the sector of charged leptons. It is interesting to investigate whether those transitions are within the present or future experimental reach. For instance, LFV decays could be large enough to be observable in supersymmetric models~\cite{Hisano:1995cp}, in the little Higgs model~\cite{Choudhury:2006sq}, and in the triplet Higgs model~\cite{Kakizaki:2003jk}. In this paper we study LFV processes in the so-called color octet model~\cite{FileviezPerez:2009ud}, which generates Majorana neutrino mass and mixing at one-loop level through the interactions of leptons with new color-octet fermions and scalars. The radiative and pure leptonic LFV decays of the muon in this model have been considered earlier in Ref.~\cite{Liao:2009fm}, neutrinoless double beta decay has been studied in Ref.~\cite{Choubey:2012ux}, and the feasibility of detecting new colored particles of TeV scale at the LHC examined in Ref.~\cite{FileviezPerez:2010ch}.

The paper is organized as follows. In the next section we introduce the color octet model and discuss the new Yukawa couplings that are most relevant to our study here. In Sec.~\ref{sec:analytic} we calculate several processes in the model: the LFV decays of the Higgs and $Z$ bosons, $h,~Z\to\ell_\alpha\overline{\ell_\beta}$ ($\ell_\alpha\ne\ell_\beta$), and the $\mu-e$ conversion in nuclei. In Sec.~\ref{sec:numerical} we illustrate our numerical results and discuss experimental constraints on the parameter space arising from the above processes together with rare muon decays $\mu\to e\gamma,~ee\bar e~(3e)$. In the last section we summarize briefly our results and conclusions. The relevant nuclear physics quantities and one-loop functions are listed in the Appendix \ref{app:A} and \ref{app:B} respectively.

\begin{table}
\centering
\begin{tabular}{ c c c c }
\hline
Type & Main subprocess & Present bound & Future sensitivity\\
\hline
\multirow{3}{*}{$\ell_\alpha\to \ell_\beta\gamma$}
&$\mu\to e\gamma$      & $4.2\times10^{-13}$ \cite{TheMEG:2016wtm}  & $\sim6\times10^{-14}$ \cite{Baldini:2013ke}\\
&$\tau\to e\gamma$      & $3.3\times10^{-8}$ \cite{Aubert:2009ag}    & $\sim3\times10^{-9}$ \cite{Aushev:2010bq}\\
&$\tau\to \mu\gamma$  & $4.4\times10^{-8}$ \cite{Aubert:2009ag}    & $\sim3\times10^{-9}$ \cite{Aushev:2010bq}\\
\hline
\multirow{5}{*}{$\ell_\alpha\to \ell_\beta\overline{\ell_\rho}\ell_\sigma$}
&$\mu\to ee^+e$                      & $1.0\times10^{-12}$ \cite{Bellgardt:1987du,Agashe:2014kda,Hayasaka:2010np} & $\sim10^{-16}$ \cite{Blondel:2013ia}\\
&$\tau\to ee^+e$                      &$ 2.7\times10^{-8}$ \cite{Agashe:2014kda,Hayasaka:2010np}   & $\sim10^{-9}$ \cite{Aushev:2010bq}\\
&$\tau\to\mu\mu^+\mu$          &$ 2.1\times10^{-8}$ \cite{Agashe:2014kda,Hayasaka:2010np}   & $\sim10^{-9}$ \cite{Aushev:2010bq}\\
&$\tau\to e\mu^+\mu$ & $2.7\times10^{-8}$ \cite{Agashe:2014kda,Hayasaka:2010np}    & $\sim10^{-9}$ \cite{Aushev:2010bq}\\
&$\tau\to e\mu^+e$ & $1.5\times10^{-8}$ \cite{Agashe:2014kda,Hayasaka:2010np}    & $\sim10^{-9}$ \cite{Aushev:2010bq}\\
&$\tau\to \mu e^+e$     & $1.8\times10^{-8}$ \cite{Agashe:2014kda,Hayasaka:2010np}    & $\sim10^{-9}$ \cite{Aushev:2010bq}\\
\hline
\multirow{4}{*}{$\mu \text{N}\to e\text{N}$}
&$\mu\text{Ti} \to e\text{Ti}$            & $4.3\times10^{-12}$ \cite{Dohmen:1993mp} & $\sim10^{-18}$ \cite{Barlow:2011zza}\\
&$\mu\text{Au} \to e\text{Au}$          & $7.0\times10^{-13}$ \cite{Bertl:2006up} &                                                  \\
&$\mu\text{Al} \to e\text{Al}$   &                                                   & $10^{-15}-10^{-18}$ \cite{Litchfield:2014qea}\\
&$\mu\text{SiC} \to e\text{SiC}$ &                                                   & $\sim10^{-14}$ \cite{H. Natori}\\
&$\mu\text{Pb}\to e\text{Pb}$ &$4.6\times 10^{-11}$ \cite{Honecker:1996zf}&\\
\hline
\multirow{3}{*}{$Z\to\ell_\alpha\overline{\ell_\beta}$}
&$Z\to\tau\mu $      & $1.2\times10^{-5}$ \cite{Abreu:1996mj} & \\
&$Z\to\tau  e $      & $2.2\times10^{-5}$ \cite{Abreu:1996mj}    & \\
&$Z\to\mu e$         & $7.3\times10^{-7}$ \cite{CMS:2015hga}  & \\
\hline
\multirow{3}{*}{$h\to\ell_\alpha\overline{\ell_\beta}$}
&$h\to\tau \mu $      & $0.84\times10^{-2}$ \cite{Khachatryan:2015kon} & \\
&$h\to\tau e $      & $7\times10^{-3}$ \cite{CMS:2015udp}    & \\
&$h\to\mu e$         & $3.6\times10^{-4}$ \cite{CMS:2015udp}  & \\
\hline
\hline
\end{tabular}
\caption{Current experimental bounds and future sensitivities on some LFV processes.}
\label{tab:experimental_data}
\end{table}

\section{Color octet model}\label{sec:model}

In the color octet model for radiative neutrino mass~\cite{FileviezPerez:2009ud}, the SM is extended by adding $N_S$ species of color octet scalars and $N_F$ species of octet fermions. The octet scalars, $S_{ar}\equiv(S_{ar}^+, S_{ar}^0)^{\T}$, have quantum numbers $(8,2,1/2)$ under the SM gauge group $SU(3)_C\otimes SU(2)_L\otimes U(1)_Y$, where $a$ denotes the color index and $r$ enumerates the species of scalars. The octet fermions have zero hypercharge but can be a singlet $\rho$ or a triplet $\chi$ under $SU(2)_L$, which are named as:
\begin{eqnarray}\label{eq:case ab}
\nonumber
\textrm{case A:}&&\rho_{ax}\sim(8,1,0),
\\
\textrm{case B:}&&
\chi_{ax}=\begin{pmatrix} \frac{1}{\sqrt{2}}\chi_{ax}^0 & \chi_{ax}^+
\\
\chi_{ax}^- & -\frac{1}{\sqrt{2}}\chi_{ax}^0
\end{pmatrix}
\sim(8,3,0),
\end{eqnarray}
where $x$ enumerates the fermions. In our discussion we focus on the scenario with two species of fermions and one scalar (i.e., $N_S=1,~N_F=2$), which is the simplest choice for generating two massive neutrinos in accord with experimental observation. From now on the scalar index $r$ is dropped while the fermion index $x$ assumes values $1,~2$.

We start with the relevant terms in the scalar potential:
\begin{eqnarray}\label{eq:potential}
V&\supset& m_S^2S_a^{\dagger}S_a+\frac{1}{2}\lambda_1(H^{\dagger}H)(S_a^{\dagger}S_a)
+ \frac{1}{2}\lambda_2(H^{\dagger}S_a)(S_a^{\dagger}H) + \frac{1}{2}\lambda_3[(H^\dagger S_a)^2
+ \text{h.c.}],
\end{eqnarray}
where $\lambda_{1,2,3}$ are real couplings. The Higgs vacuum expectation value, $\langle H^0\rangle=v/\sqrt{2}$, causes a mass splitting among the members of the scalar doublet. Decomposing the neutral member into real and imaginary parts, $S_a^0=(S_a^R+iS_a^I)/\sqrt{2}$, the tree level mass spectrum is,
\begin{eqnarray}\label{eq:mass_splitting}
\nonumber
m_{S^\pm}^2&=&m^2_S+\frac{1}{4}\lambda_1v^2,
\\
\nonumber
m_{S^R}^2&=&m^2_S+\frac{1}{4}(\lambda_1+\lambda_2+2\lambda_3)v^2,
\\
m_{S^I}^2&=&m^2_S+\frac{1}{4}(\lambda_1+\lambda_2-2\lambda_3)v^2.
\end{eqnarray}
In this paper we will focus on color octet scalars with masses of TeV scale. The above mass splittings are expected to be smaller, and thus whenever possible, are neglected. In this case we denote the scalar mass generically by $m_S$. Furthermore, since the mass splitting between the neutral and charged members of a triplet fermion in case B is generated at one loop~\cite{FileviezPerez:2009ud} and can thus be ignored as well, we denote the fermion masses simply by $m_x$. There are some experimental constraints on those masses. The CMS Collaboration has excluded $m_S<625~\GeV$ at $95\%~\textrm{C.L.}$ in direct searches for $S$ pair production in the $Z$-gluon-$b\bar{b}$ final state~\cite{Khachatryan:2015jea}, while the ATLAS search for four tops \cite{Aad:2015gdg} and the CMS search for four jets \cite{Khachatryan:2014lpa} have excluded $m_S<830~\GeV$ at $95\%~\textrm{C.L.}$. For the octet fermions, the recent results from LHC at $13~\TeV$ in searches for supersymmetry particles like gluinos have extended the lower bound on the colored fermions up to $1.6-1.8\text{TeV}$ \cite{Khachatryan:2016kdk}.

The Yukawa couplings in SM and the additional terms in case A and B of the octet model are
\begin{eqnarray}\label{eq:Yukawa}
\nonumber
-\mathcal{L}_{\text{SM}}^{\text{Yuk}}&=&
g_{\alpha \beta}^L\overline{L_{L\alpha}}Hl_{R\beta}
+g^{U}_{\alpha\beta}\overline{Q_{L\alpha}}\tilde{H}u_{R\beta}
+g^{D}_{\alpha\beta}\overline{Q_{L\alpha}}Hd_{R\beta}+\text{h.c.},
\\
\nonumber
-\mathcal{L}_{\text{A}}^{\text{Yuk}}&=&
z_{\alpha x}\overline{L_{L\alpha}}\tilde{S_a}\rho_{ax}
+\eta_{U}g^{U}_{\alpha\beta}\overline{Q_{L\alpha}}\tilde{S_a}T_au_{R\beta}
+\eta_{D}g^{D}_{\alpha\beta}\overline{Q_{L\alpha}}S_aT_ad_{R\beta}+\text{h.c.},
\\
-\mathcal{L}_{\text{B}}^{\text{Yuk}}&=&
z_{\alpha x}\overline{L_{L\alpha}}\chi_{ax}\tilde{S_a}
+\eta_{U}g^{U}_{\alpha\beta}\overline{Q_{L\alpha}}\tilde{S_a}T_au_{R\beta}
+\eta_{D}g^{D}_{\alpha\beta}\overline{Q_{L\alpha}}S_aT_ad_{R\beta}+\text{h.c.},
\end{eqnarray}
where $L_L,~Q_L$ are the left-handed lepton and quark doublets, $l_R,~u_R,~d_R$ the right-handed singlets, and $T_a$ are $SU(3)_C$ generators in the fundamental representation. We have made the assumption of minimal flavor violation in the Yukawa couplings between quarks and the octet scalar, so that $\eta_U,~\eta_D$ are simple complex numbers \cite{Manohar:2006ga}.

\begin{figure}[!htbp]
\centering
\includegraphics[scale=2]{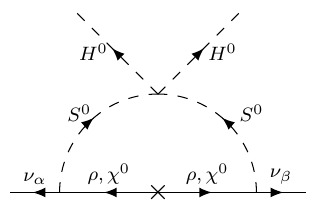}
\caption[]{Neutrino mass generated at the one-loop level in the color octet model.}
\label{fig1}
\end{figure}

The neutrino mass is generated at one loop via the Feynman graph in Fig. \ref{fig1}. We discuss case A for the purpose of illustration, for which the neutrino mass matrix reads
\begin{equation}
\label{massmatrix}
M_{\alpha\beta}=z_{\alpha x}z_{\beta x}\frac{\lambda_3 v^2}{4\pi^2}I(m_{\rho_x},m_S),
\end{equation}
where the loop integration function $I(m_{\rho_x},m_S)$ is given by
\begin{eqnarray}\label{eq:loop_factor_I}
I(m_{x},m_S)&=&\frac{m_x}{(m_S^2-m_x^2)^2}\left[m_S^2-m_x^2+m_x^2\ln\frac{m_x^2}{m_S^2}\right], \end{eqnarray}
and will be shortened as $I_x$. In the basis where the charged leptons have been diagonalized, the above neutrino mass matrix is diagonalized by the Pontecorvo-Maki-Nakagawa-Sakata (PMNS) matrix $U$: $U^\dagger M U^*=m_\nu=\text{diag}(m_{\nu_1},m_{\nu_2},m_{\nu_3})$, with $m_{\nu_{1,2,3}}$ being the neutrino masses. Since $M_{\alpha\beta}$ is degenerate \cite{Liao:2009fm} with our minimal choice of the octet species, the lightest neutrino is massless in either normal (NH) or inverted hierarchy (IH):
\begin{eqnarray}
\nonumber
\textrm{NH:}&&(m_{\nu_1},m_{\nu_2},m_{\nu_3})
=\left(0,\sqrt{\Delta m_{21}^2},\sqrt{\Delta m_{31}^2}\right),
\\
\textrm{IH:}&&(m_{\nu_1},m_{\nu_2},m_{\nu_3})
=\left(\sqrt{|\Delta m_{31}^2|},\sqrt{|\Delta m_{31}^2|+\Delta m_{21}^2},0\right).
\label{eq:neutrino_mass}
\end{eqnarray}
The global fit in Ref.~\cite{Capozzi:2013csa} yields the following best-fit values for the mass splittings $\Delta m_{ij}^2=m_{\nu_i}^2-m_{\nu_j}^2$, the mixing angles $\theta_{ij}$, and the Dirac CP phase $\delta$:
\begin{eqnarray}\label{eq:neutrino_data}
\nonumber
&&\Delta m_{21}^2=7.54\times 10^{-5}\text{eV}^2, \  \Delta m_{31}^2=2.47~(-2.34)\times 10^{-3}\text{eV}^2,
\\
\nonumber
&&\sin^2\theta_{12}=3.08\times10^{-1},  \  \sin^2\theta_{13}=2.34~(2.40)\times10^{-2},
\\
&&\sin^2\theta_{23}=4.37~(4.55)\times10^{-1}, \  \delta=1.39~(1.31)\pi,
\end{eqnarray}
where the number in parentheses refers to IH when it differs from the NH case.

The special structure of Eq. \eqref{massmatrix} allows to solve the Yukawa couplings $z$ in terms of the neutrino masses $m_{\nu}$, mixing matrix $U$ and a free complex number $\omega$  \cite{Casas:2001sr}:
\begin{equation}\label{eq:para}
z=\sqrt{\frac{4\pi^2}{\lambda_3}}\frac{1}{v}U(m_\nu)^{1/2}\Omega
\begin{pmatrix}
  I_1  &  \\
  &   I_2
\end{pmatrix}^{-1/2},
\end{equation}
where for NH and IH cases one has respectively,
\begin{eqnarray}
\Omega_{\NH}=
\begin{pmatrix}
0 & 0 \\
\sqrt{1-\omega^2} & -\omega \\
\omega & \sqrt{1-\omega^2}\\
\end{pmatrix},
\quad
\Omega_{\IH}=
\begin{pmatrix}
\sqrt{1-\omega^2} & -\omega \\
\omega & \sqrt{1-\omega^2}  \\
0 &0 \\
\end{pmatrix}.
\end{eqnarray}
Some comments are in order. The existence of two massive neutrinos requires the two octet fermions to be nondegenerate, because if they are degenerate only a linear combination of them couples to the leptons so that the Yukawa couplings $z$ effectively become a column matrix and only one neutrino can gain mass at one loop. In our numerical analysis, we will employ Eqs. (\ref{eq:neutrino_mass},\ref{eq:neutrino_data}) in Eq. (\ref{eq:para}) but ignore in $I_{1,2}$ the mass splitting between the two octet fermions. This should be taken as a technical simplification to reduce free parameters instead of any inconsistency. In some of our numerical illustrations we will restrict ourselves to the case of a pure phase $\omega=\exp(i2\pi\kappa)$ with $\kappa\in[0,1]$, while for other numerical analyses we will consider a real $\omega\in[-1,1]$. In the latter case, our key parameter $z_{ex}z_{\mu x}^*$, where $x$ is summed over, becomes independent of the real $\omega$ parameter when the mass splitting is ignored in $I_{1,2}\approx I_0$, e.g., in the $e\mu$ sector:
\begin{eqnarray}
\textrm{NH:}&&z_{ex}z_{\mu x}^*
=\frac{4\pi^2}{v^2\lambda_3I_0}(U_{e2}U_{\mu 2}^*m_{\nu_2}+U_{e3}U_{\mu 3}^*m_{\nu_3}),
\nonumber
\\
\textrm{IH:}&&z_{ex}z_{\mu x}^*
=\frac{4\pi^2}{v^2\lambda_3I_0}(U_{e1}U_{\mu 1}^*m_{\nu_1}+U_{e2}U_{\mu 2}^*m_{\nu_2}),
\label{eq:common_factor1}
\end{eqnarray}
and similarly for general $z_{\alpha x}z_{\beta x}^*$.

%%%%%%%%%%%%
\section{Analytic results}\label{sec:analytic}
%%%%%%%%%%%%

In this section we will present our analytic results for the three types of processes, $\mu e$ conversion in nuclei, $h \to\ell_\alpha \overline{\ell_\beta}$ and $Z \to\ell_\alpha \overline{\ell_\beta}$. We will ignore the tiny SM contributions from the start.

%%%%%%%%%%%%
\subsection{$h\to\ell_\alpha \overline{\ell_\beta}$}
%%%%%%%%%%%%

\begin{figure}[!htbp]
\centering
\includegraphics[scale=1.2]{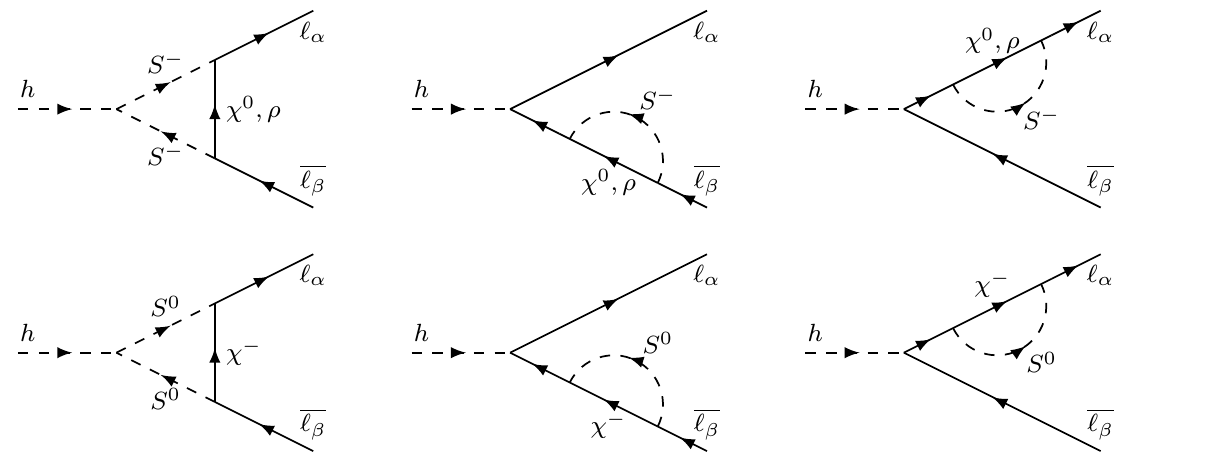}
\caption[]{Feynman diagrams for LFV Higgs decays due to octet particles.}
\label{fig2}
\end{figure}

The Feynman graphs for the LFV decays of the Higgs boson, $h\to \ell_\alpha \overline{\ell_\beta}$ ($\ell_\alpha\ne\ell_\beta$), are shown in Fig. \ref{fig2}. We have dropped the terms proportional to the small ratio $m^2_h/m^2_S$, and similarly we will drop $m^2_Z/m^2_S$ terms for the LFV decays of the $Z$ boson. The amplitude is,
\begin{eqnarray}\label{hLFVamp}
\nonumber
\mathcal{M}(h \to\ell_\alpha \overline{\ell_\beta})&=&
\frac{Cz_{\alpha x}z^*_{\beta x}v}{16\pi^2m^2_S}
\Big\{\left[\xi\lambda_1+2(1-\xi)(\lambda_1+\lambda_2)\right]F(r_x)
\overline{u_\alpha}(m_\alpha P_L+m_\beta P_R)v_\beta
\\
&&+2(2-\xi)v^{-2}m_\alpha m_\beta F_2(r_x)
\overline{u_\alpha}(m_\alpha P_R+m_\beta P_L)v_\beta\Big\},
\end{eqnarray}
where $\xi=1~(1/2)$ for case A (B), $C=8$ counts the color number of new particles, and $m_{\alpha,\beta}$ are the lepton masses. The loop functions $F$ and $F_2$ of the fermion to scalar mass ratios $r_x=m^2_{x}/m^2_S$ are listed in Appendix \ref{app:B}. The branching ratio is found to be, assuming $m_\beta\gg m_\alpha$,
\begin{eqnarray}\label{hLFVBr}
\nonumber
{\Br}(h \to\ell_\alpha \overline{\ell_\beta})&=&
\frac{m_h}{\Gamma_{h}}\frac{m^2_\beta v^2}{2^7\pi^5m^4_{S}}
%\\
%\nonumber
%&&
\Big\{\Big|z_{\alpha x}z^*_{\beta x}
\Big(\left[\xi\lambda_1+2(1-\xi)(\lambda_1+\lambda_2)\right]F(r_x)
+2(2-\xi)v^{-2}m_\alpha m_\beta F_2(r_x)\Big)\Big|^2
\\
&&+\Big|z_{\alpha x}z^*_{\beta x}
\Big(\left[\xi\lambda_1+2(1-\xi)(\lambda_1+\lambda_2)\right]F(r_x)
-2(2-\xi)v^{-2}m_\alpha m_\beta F_2(r_x)\Big)\Big|^2\Big\},
\end{eqnarray}
where $\Gamma_{h}\approx 5~\MeV$ is the Higgs total decay width \cite{Agashe:2014kda}. It is clear that the branching ratio is severely suppressed by the heavy masses of the octet particles.

%%%%%%%%%%%%
\subsection{$Z\to\ell_\alpha \overline{\ell_\beta}$}
%%%%%%%%%%%%

\begin{figure}[!htbp]
\centering
\includegraphics[scale=1.2]{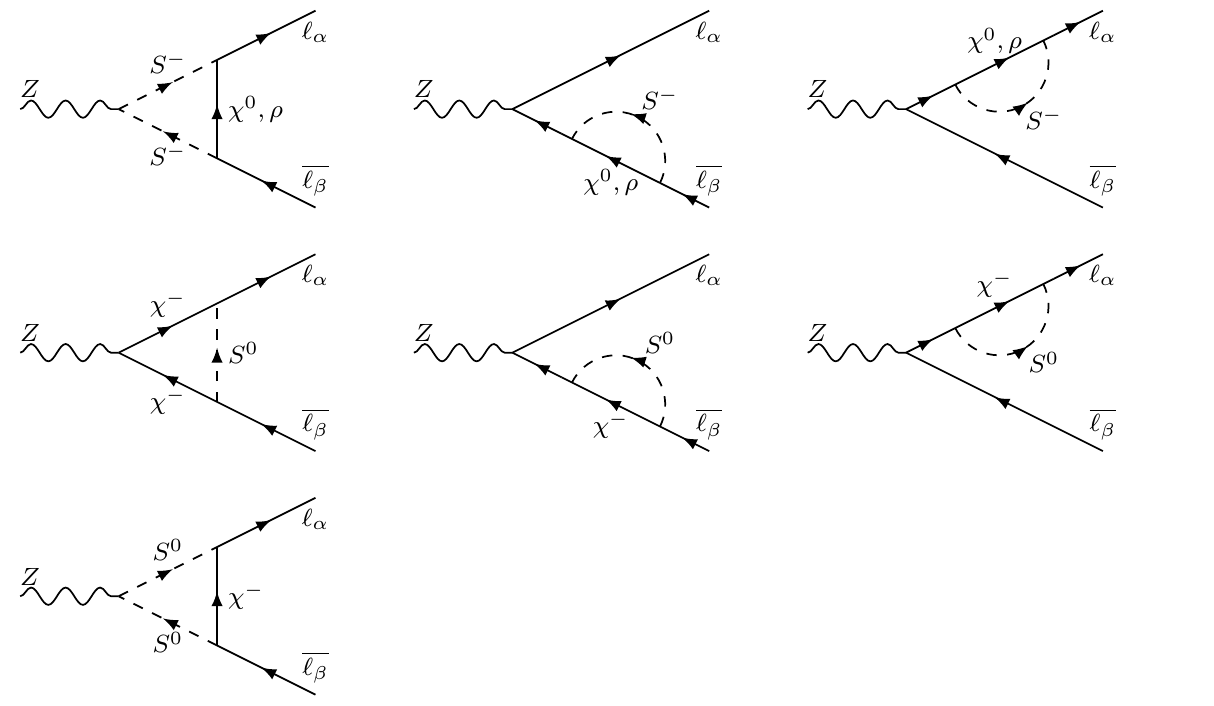}
\caption[z_decay_feyn]{Feynman diagrams for LFV $Z$ decays due to octet particles.}
\label{fig3}
\end{figure}

The Feynman diagrams for the LFV decays of the $Z$ boson are shown in Fig. \ref{fig3}. Compared with the LFV decays of the Higgs boson there is an additional diagram in case B (with a triplet octet fermion), which is essential to make the whole amplitude free of UV divergence. The amplitude is,
\begin{eqnarray}\label{ZLFVamp}
\nonumber
\mathcal{M}(Z \to\ell_\alpha \overline{\ell_\beta})&=&
\frac{Cg_2}{32\pi^2 \cos\theta_W m^2_S} \epsilon^\mu
\overline{u_\alpha}\Big\{C_{\alpha\beta}^Z(k^2\gamma_\mu - k_\mu \slashed{k})P_L
- D_{\alpha\beta}^Z(m_\alpha P_L + m_\beta P_R) i\sigma_{\mu\nu} k^\nu
\\
&&+E_{\alpha\beta}^Z(m_\alpha^2 + m_\beta^2)\gamma_\mu P_L  +  F_{\alpha\beta}^Zm_\alpha m_\beta\gamma_\mu P_R \Big\}v_\beta,
\end{eqnarray}
where $\theta_W$ is the weak mixing angle, $g_2$ the $SU(2)_L$ gauge coupling, and $k$ and $\epsilon$ are the momentum and polarization of the $Z$ boson. The above effective interaction will enter the $\mu-e$ conversion in nuclei, and its form factors $C_{\alpha\beta}^Z,~D_{\alpha\beta}^Z,~E_{\alpha\beta}^Z,~F_{\alpha\beta}^Z$ are given in Eq. (\ref{eq:A}). Dropping the terms suppressed by the lepton masses, the branching fraction is found to be,
\begin{eqnarray}\label{eq:z_branching}
\Br(Z\to\ell_\alpha\overline{\ell_\beta})&=&
\frac{1}{\Gamma_{Z}}\frac{G_Fm^7_Z }{3\sqrt{2}\cdot 2^{5}\pi^5 m^4_S}|C_{\alpha\beta}^Z|^2,
\end{eqnarray}
where $\Gamma_{Z}=2.4952~\GeV$ is the total decay width of the $Z$ boson \cite{Agashe:2014kda}.

%%%%%%%%%%%%
\subsection{$\mu N\to eN$}\label{mue}
%%%%%%%%%%%%

\begin{figure}[!htbp]
\centering
\includegraphics[scale=1.1]{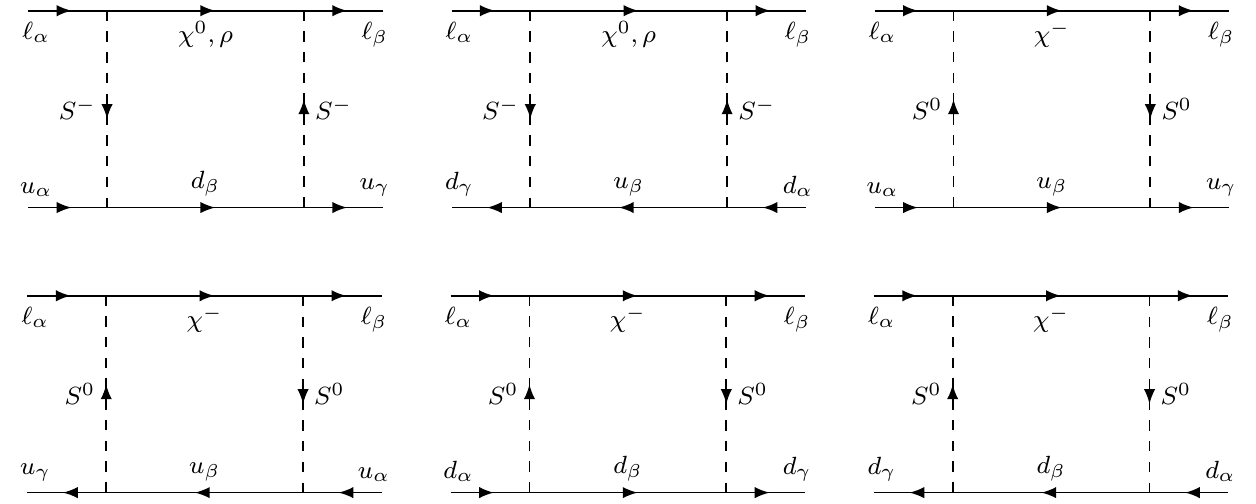}\\
\vspace{0.9cm}
\includegraphics[scale=1.1]{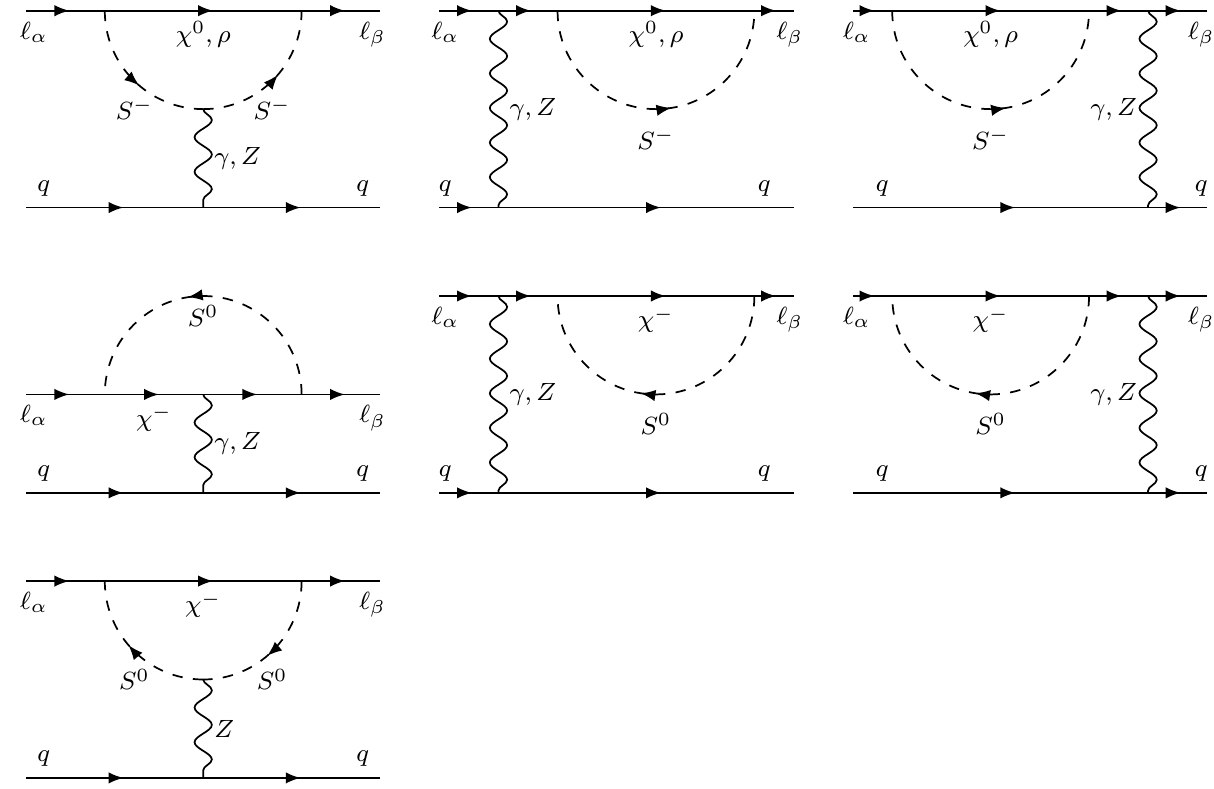}
\caption[]{Feynman diagrams relevant for $\mu-e$ conversion in nuclei due to octet particles.}
\label{fig4}
\end{figure}

The LFV decays of the Higgs and $Z$ bosons can only be studied at high energy colliders, and their current experimental limits are rather weak. The most dramatic experimental advances concerning LFV processes in the near future are expected to take place in the LFV decays of the muon and $\mu-e$ conversion in atomic nuclei. In the color octet model, the radiative and pure leptonic decays of the muon have been studied in Ref.~\cite{Liao:2009fm}. In this work, we concentrate on the coherent $\mu-e$ conversion in nuclei, which is generally much more significant than its incoherent counterpart~\cite{Kitano:2002mt}.

The most general Lagrangian at the quark level that is relevant to $\mu-e$ conversion in nuclei can be parameterized as follows~\cite{Kitano:2002mt}:
\begin{eqnarray}\label{g_L}
\nonumber
\mathcal{L}_\textrm{int}&\supset &
-\frac{4G_F}{\sqrt{2}}m_\mu \left(A_R\overline{\mu}\sigma^{\mu\nu}P_LeF_{\mu\nu}
+ A_L\overline{\mu}\sigma^{\mu\nu}P_ReF_{\mu\nu}+\text{h.c.}\right)
\\
&&-\frac{G_F}{\sqrt{2}}\sum_{q=u,d,s}\Big\{
\left(g_{LS}^q\overline{e}P_R\mu+g_{RS}^q\overline{e}P_L\mu\right)\overline{q}q
+\left(g_{LV}^q\overline{e}\gamma_\mu P_L\mu+g_{RV}^q\overline{e}\gamma_\mu P_R\mu\right)\overline{q}\gamma^\mu q+\text{h.c.}\Big \},
\end{eqnarray}
where we have neglected the pseudoscalar and axial vector currents of quarks as they have no contributions to the coherent $\mu-e$ conversion. $A_{L,R}$ and various $g^q$ are dimensionless effective couplings. The branching ratio for the conversion can be written as:
\begin{eqnarray}\label{eq:branching}
\nonumber
\Br(\mu N \to eN)
&=&2G_F^2\Big(|A_R^*D+g_{LS}^pS^p+g_{LS}^nS^n+g_{LV}^pV^p+g_{LV}^nV^n|^2
\\
&&+|A_L^*D+g_{RS}^pS^p+g_{RS}^nS^n+g_{RV}^pV^p+g_{RV}^nV^n|^2\Big)\Gamma_{\capt}^{-1},
\end{eqnarray}
where $\Gamma_{\capt}$ is the $\mu$ capture rate in the atomic nucleus. The effective couplings $g^{p(n)}$ for the proton (neutron) in Eq. \eqref{eq:branching} are built from those of quarks in Eq. (\ref{g_L}) by
\begin{eqnarray}
\label{eq:coefficients}
&&g_a^{p(n)}=\sum_qG_S^{p(n),q}g_a^q~(a=LS,~RS),\quad
g_b^{p(n)}=\sum_qG_V^{p(n),q}g_b^q~(b=LV,~RV).
\end{eqnarray}
The values of the coefficients $G_{S,V}^{p(n),q}$, $\Gamma_{\capt}$, and the overlap integrals $D,~S^{p(n)},~V^{p(n)}$ for various nuclei can be extracted from Ref.~\cite{Kitano:2002mt} and are reproduced in Appendix \ref{app:A}.

In the color octet model, the $\mu-e$ conversion arises at the one-loop level and the Feynman diagrams can be divided into three classes: the $\gamma$ penguin, the $Z$ penguin, and the box diagrams as shown in Fig. \ref{fig4}. We have neglected the Higgs penguin contribution as it is heavily suppressed by the light quark Yukawa couplings. The amplitude for the photonic transition $\ell_\alpha\to\ell_\beta\gamma^{(*)}(k)$ expanded to the first nontrivial order in external momenta is~\cite{Liao:2009fm},
\begin{eqnarray}\label{lepton_radiative_decays}
\mathcal{M_\mu}&=&-\frac{Ce}{(4\pi)^2m_S^2}\overline{u_\beta}
\left\{A_{\beta\alpha}^\gamma(k^2\gamma_\mu-k_\mu \slashed{k})P_L
+ B_{\beta\alpha}^\gamma(m_\alpha P_R + m_\beta P_L) i\sigma_{\mu\nu} k^\nu \right\}u_\alpha.
\end{eqnarray}
While the dipole term is already in the form of Eq. (\ref{g_L}), the anapole term can be converted to the vector-vector form when the photon is connected to a quark. Incorporating the latter (first term in Eq. \ref{effective_L}) in the non-photonic contributions from the $Z$ penguin and box diagrams yields the following terms for the $\ell_\alpha \to\ell_\beta$ conversion in nuclei:
\begin{eqnarray}\label{effective_L}
\nonumber
\mathcal{M}'&=&
-\frac{C\alpha}{4\pi m_S^2}A_{\beta\alpha}^\gamma
\overline{\ell_\beta}\gamma_\mu P_L \ell_\alpha \sum_{q=u,d,s}Q^q\overline{q}\gamma^\mu q
\\
\nonumber
&&-\frac{G_F}{\sqrt{2}}\frac{C}{4\pi^2m_S^2}\overline{\ell_\beta} \Big\{C_{\beta\alpha}^Z(k^2\gamma_\mu - k_\mu \slashed{k})P_L
+  D_{\beta\alpha}^Z(m_\alpha P_R + m_\beta P_L) i\sigma_{\mu\nu} k^\nu
\\
\nonumber
&&+E_{\beta\alpha}^Z(m_\alpha^2 + m_\beta^2)\gamma_\mu P_L
+  F_{\beta\alpha}^Zm_\alpha m_\beta\gamma_\mu P_R \Big\} \ell_\alpha \sum_{q=u,d,s}\frac{1}{2}(Z_L^q + Z_R^q)\overline{q}\gamma^\mu q
\\
\nonumber
&&-\frac{G_F}{\sqrt{2}}\frac{-C_F}{8\pi^2m_S^2}G_{\beta\alpha}^\textrm{box}
\Big\{\sum_{q=u}(\eta_U\eta_U^*m_{q}^2
+ \eta_D\eta_D^*V_{qq^\prime}m_{q^\prime}^2V_{q^\prime q}^\dagger)
\overline{\ell_\beta}\gamma_\mu P_L \ell_\alpha \overline{q}\gamma^\mu q
\\
&&-\sum_{q=d,s}(\eta_D\eta_D^*m_{q}^2
+\eta_U\eta_U^*V_{qq^\prime}^\dagger m_{q^\prime}^2V_{q^\prime q})
\overline{\ell_\beta}\gamma_\mu P_L\ell_\alpha \overline{q}\gamma^\mu q\Big\},
\end{eqnarray}
where $C_F=4/3$, $Z_{L/R}^q= T^{3,q}_{L/R}-Q^q\sin^2\theta_W$, $Q^q$ the charge of quark $q$ in units of $|e|$, $V$ the CKM matrix, and $k$ is the virtual $Z$ momentum from lepton to quark lines. We have neglected axial vector quark currents. The coefficients are found to be
\begin{eqnarray}\label{eq:A}
\nonumber
A_{\beta\alpha}^\gamma&=&z_{\beta x}z_{\alpha x}^*\left[\xi F_1(r_x)+2(1-\xi)G_1(r_x)\right],
\\
\nonumber
B_{\beta\alpha}^\gamma&=&z_{\beta x}z_{\alpha x}^*\left[\xi F_2(r_x)+2(1-\xi)G_2(r_x)\right],
\\
\nonumber
C_{\beta\alpha}^Z&=&z_{\beta x}z_{\alpha x}^*
\left\{\left[2(1- \xi) - \xi\cos(2\theta_W)\right]F_1(r_x)
- 4(1-\xi)\cos^2\theta_W G_1(r_x)\right\},
\\
\nonumber
D_{\beta\alpha}^Z&=&z_{\beta x}z_{\alpha x}^*
\left\{\left[2(1\!-\!\xi)-\xi\cos(2\theta_W)\right]F_2(r_x)
- 4(1-\xi)\cos^2\theta_WG_2(r_x)\right\},
\\
\nonumber
E_{\beta\alpha}^Z&=&-z_{\beta x}z_{\alpha x}^*2(1-\xi)F_2(r_x),
\\
\nonumber
F_{\beta\alpha}^Z&=&-z_{\beta x}z_{\alpha x}^*2(2-\xi)F_2(r_x),
\\
G_{\beta\alpha}^\textrm{box}&=&z_{\beta x}z_{\alpha x}^*\xi H(r_x),
\end{eqnarray}
where summation over the octet fermion species $x$ is implied and the loop functions $F_{1,2}(x),~G_{1,2}(x),~H(x)$ are listed in Appendix \ref{app:B}. Since the $Z$ penguin and box diagrams are suppressed by lepton and light quark masses, their contributions can actually be neglected in our numerical analysis. But we should be aware that which contribution dominates can be model dependent; for a model-independent analysis on $\mu-e$ conversion, one can see, e.g., Ref.~\cite{Cirigliano:2009bz}. From now on, we keep only the photonic contribution and suppress its label from the relevant coefficients. Comparing Eqs.~(\ref{lepton_radiative_decays},\ref{effective_L}) with \eqref{g_L}, we finally obtain the form factors in Eq. \eqref{eq:branching}:
\begin{eqnarray}
\nonumber
A_{R,L}^*&=&
\frac{-\sqrt{2}Ce}{128\pi^2G_Fm_S^2}\big(1,m_e/m_\mu\big)B_{e\mu},
\\
g_{LV}^{p}&=&
\frac{\sqrt{2}C\alpha}{4\pi G_F m_S^2}A_{e\mu} \sum_{q=u,d,s}G_V^{p,q}Q^q,
\label{eq:form_factors}
\end{eqnarray}
where $A_L^*$ can be ignored comparing with $A_R^*$. Combining Eqs. (\ref{eq:branching},\ref{eq:form_factors}) yields a simple branching ratio:
\begin{eqnarray}\label{eq:simplified_branching}
\text{Br}(\mu N\to eN)&=&
\frac{\alpha}{16\pi^3}\frac{|B_{e\mu}D-16\sqrt{\pi\alpha}A_{e\mu}V^p|^2}{\Gamma_\textrm{capt}m_S^4}.
\end{eqnarray}

%%%%%%%%%%%%%%
\section{Numerical analysis}\label{sec:numerical}
%%%%%%%%%%%%%%

%%%%%%%%%%%%%%
\subsection{$h \to\ell_\alpha \overline{\ell_\beta}$ and $Z \to\ell_\alpha \overline{\ell_\beta}$ }
%%%%%%%%%%%%%%

\begin{figure}[!htbp]
\centering
\includegraphics[scale=0.5]{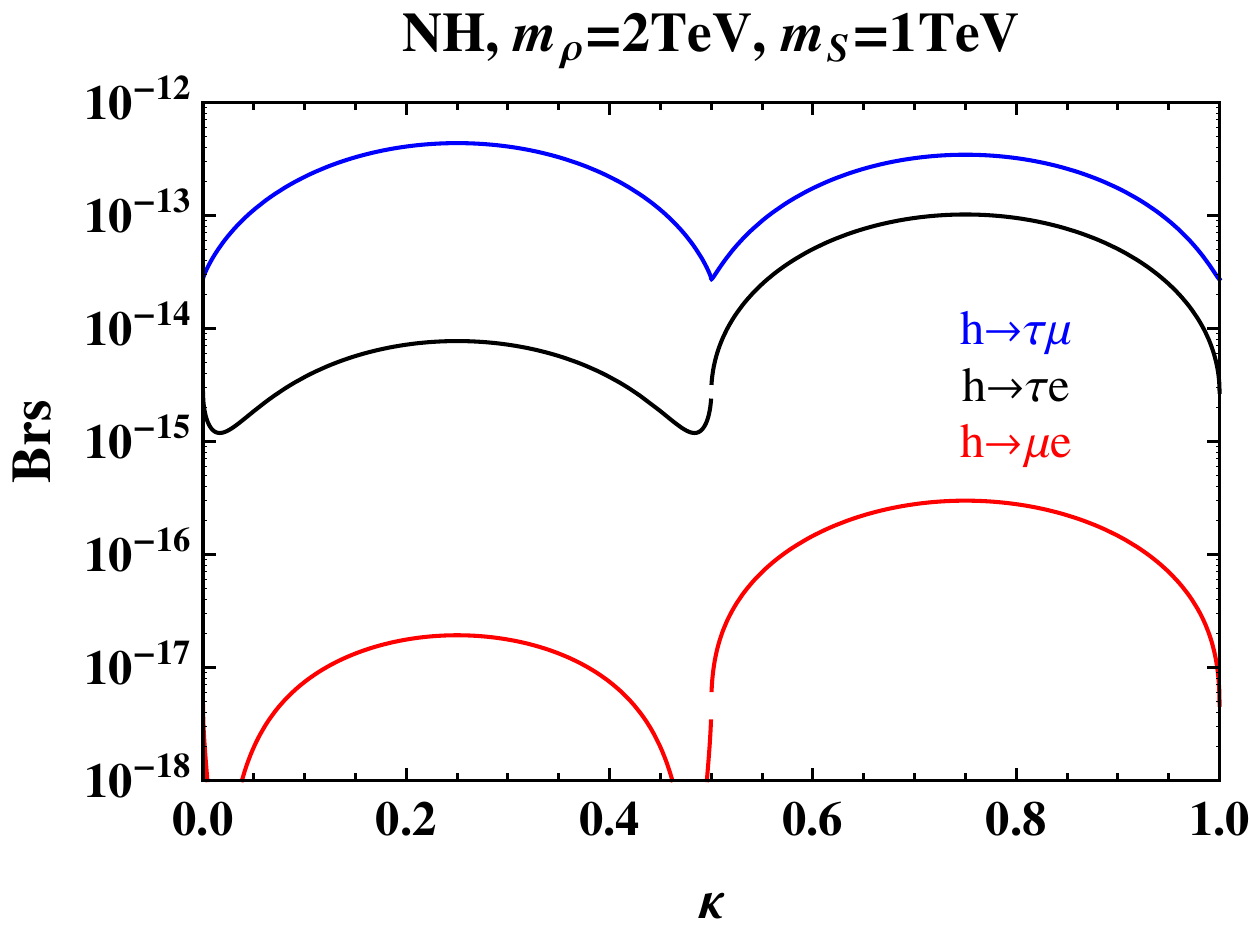}
\quad
\includegraphics[scale=0.5]{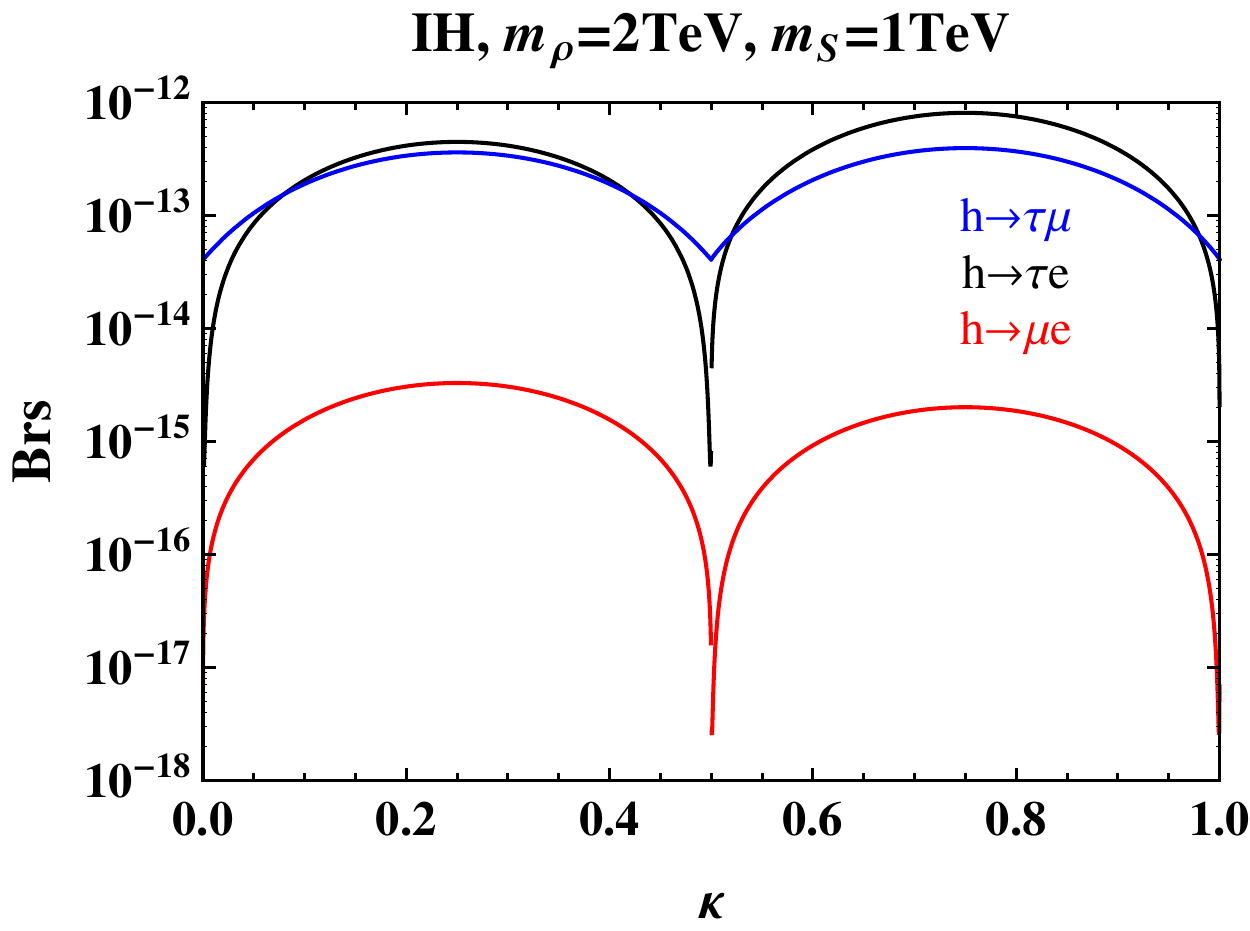}\\
\vspace{5pt}
\includegraphics[scale=0.5]{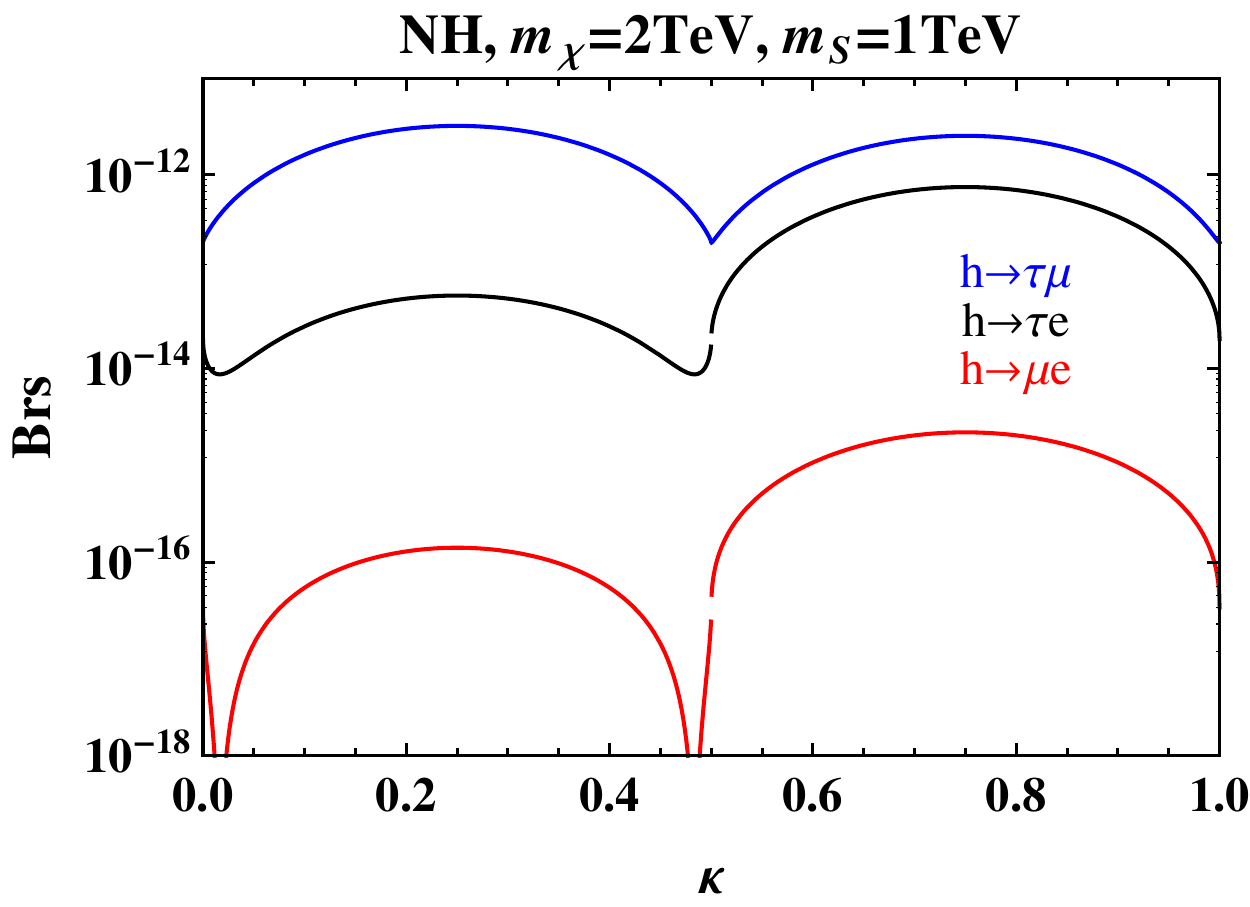}
\quad
\includegraphics[scale=0.5]{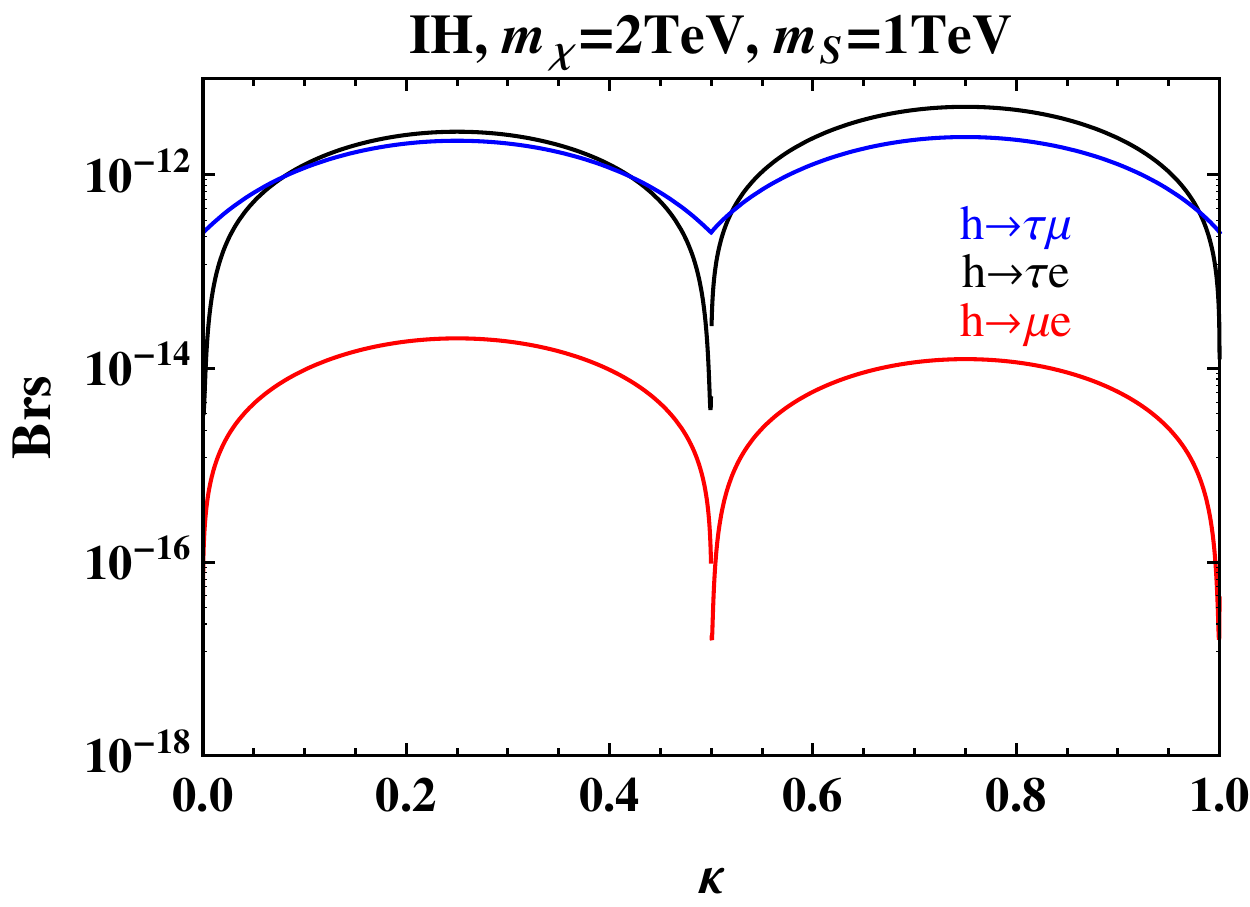}
\caption{Branching fractions of LFV Higgs decays are shown as a function of $\kappa$ for both neutrino mass hierarchies (left panel for NH and right for IH) and in both case A (upper panel) and case B (lower panel).}
\label{fig5}
\end{figure}

As one can see from Eq. (\ref{hLFVBr}), the LFV Higgs decays discriminate between the two cases of singlet (case A) and triplet (case B) octet fermions through the last three graphs in Fig. \ref{fig2} that introduce the $\lambda_2$ dependence in the latter case. In Fig.~\ref{fig5} we plot the branching fractions of the LFV Higgs decays as a function of the free phase parameter $\kappa$ for two neutrino mass hierarchies (NH and IH) and in both cases A and B. We have set $\lambda_1=\lambda_2=1$, and assumed $m_{\rho(\chi)}=2~\TeV,~m_S=1~\TeV$ which are above the current experimental limits. We have following observations:
\begin{itemize}
\item All three decay channels, $h\to\tau \mu,~\tau e,~\mu e$, have a much smaller branching fraction than the current experimental bound albeit well above the SM expectations.

\item Case B yields one order of magnitude enhancement compared to case A due to the involvement of more colored particles.

\item There exists a cancellation at $\kappa\approx 0.02,~0.48$ ($0.5,~1$) in NH (IH), and the cancellation is sharper in the IH case. This feature is controlled by the key parameter $z_{ex}z_{\mu x}^*$.

\item The branching fractions for NH roughly follow the hierarchy in the current experimental upper bounds, $\Br(h\to\mu \tau)>\Br(h\to e \tau)>\Br(h\to\mu e)$, but there is no similar relation for IH.
\end{itemize}

\begin{figure}[!htbp]
\centering
\includegraphics[scale=0.5]{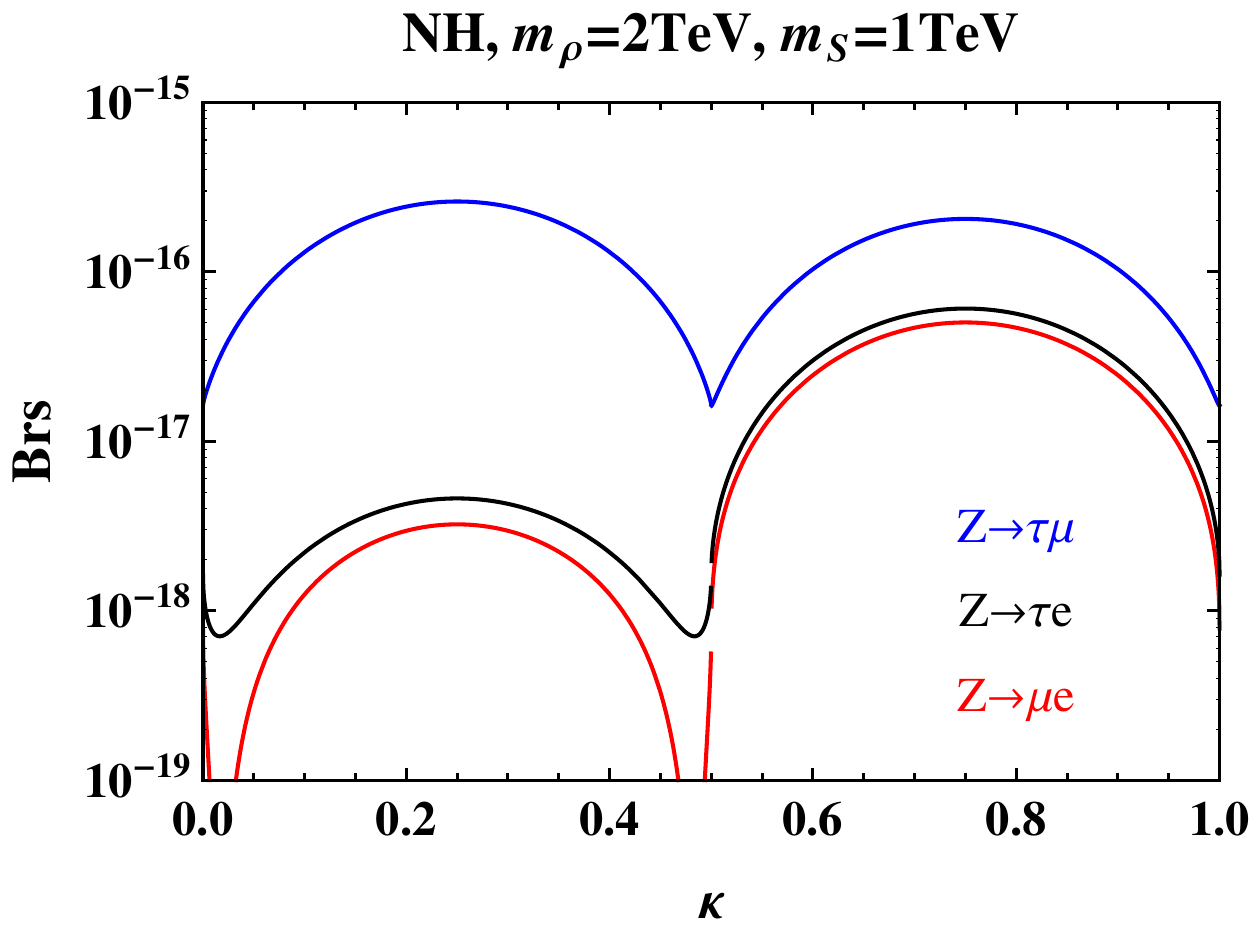}
\quad
\includegraphics[scale=0.5]{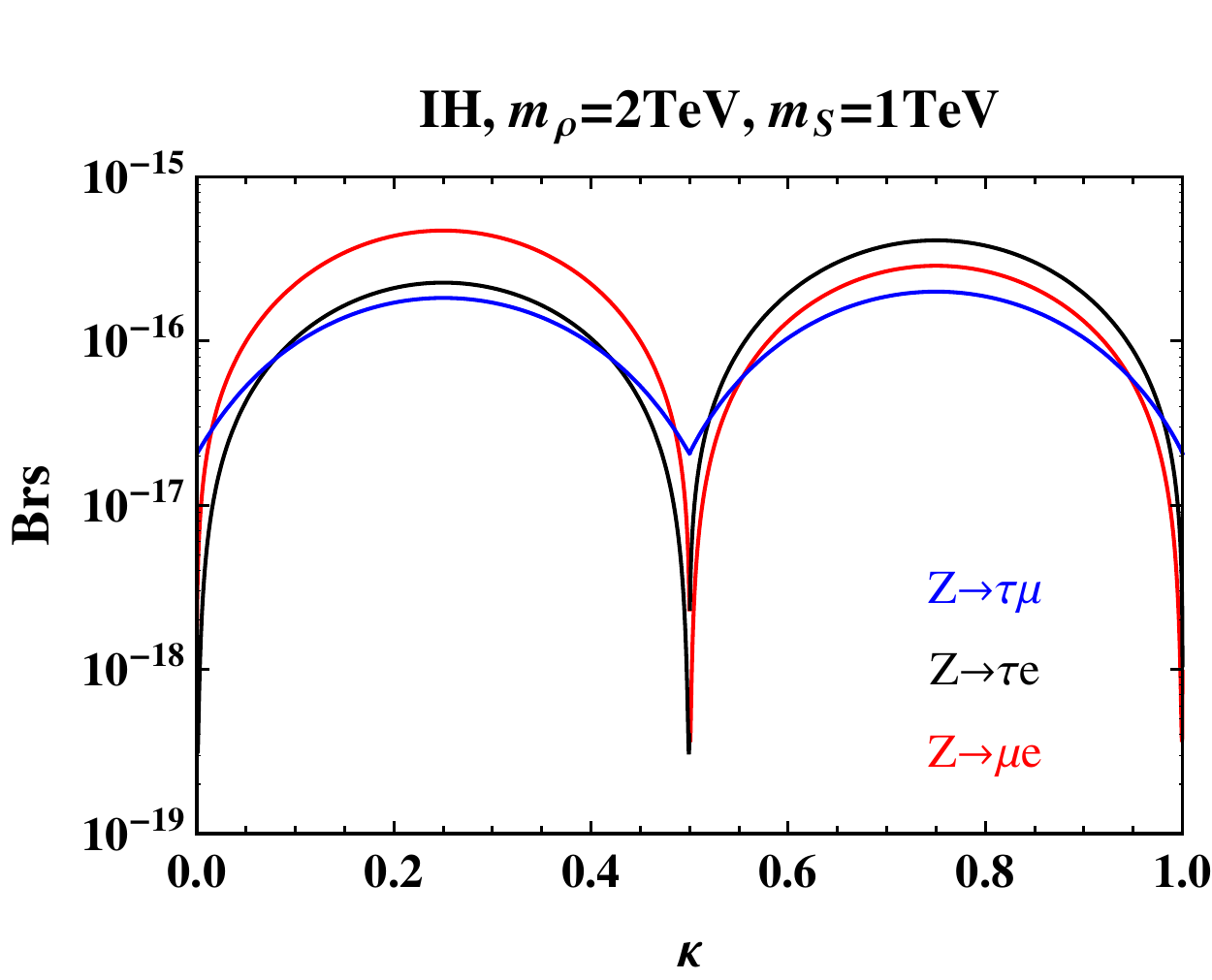}\\
\vspace{5pt}
\includegraphics[scale=0.5]{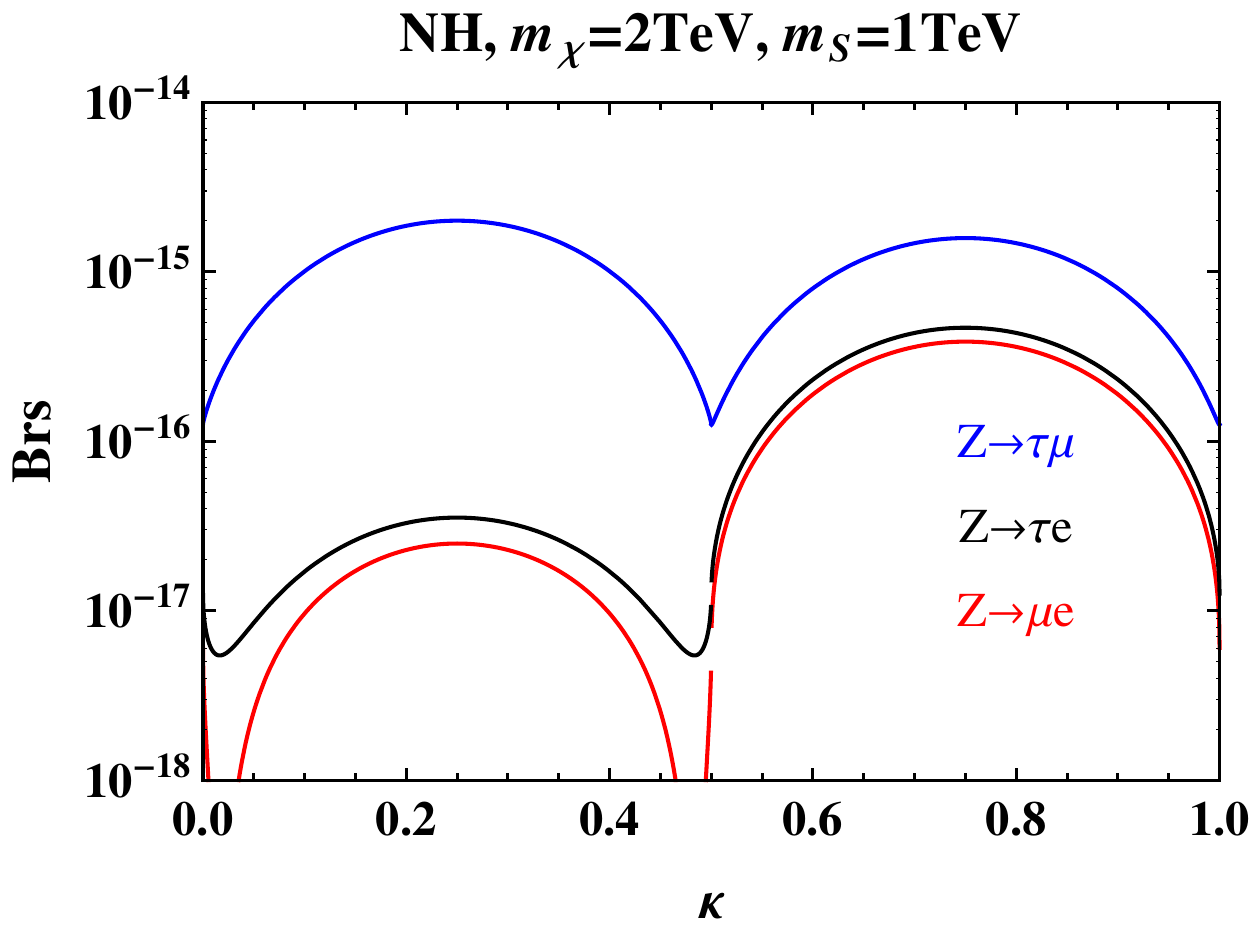}
\quad
\includegraphics[scale=0.5]{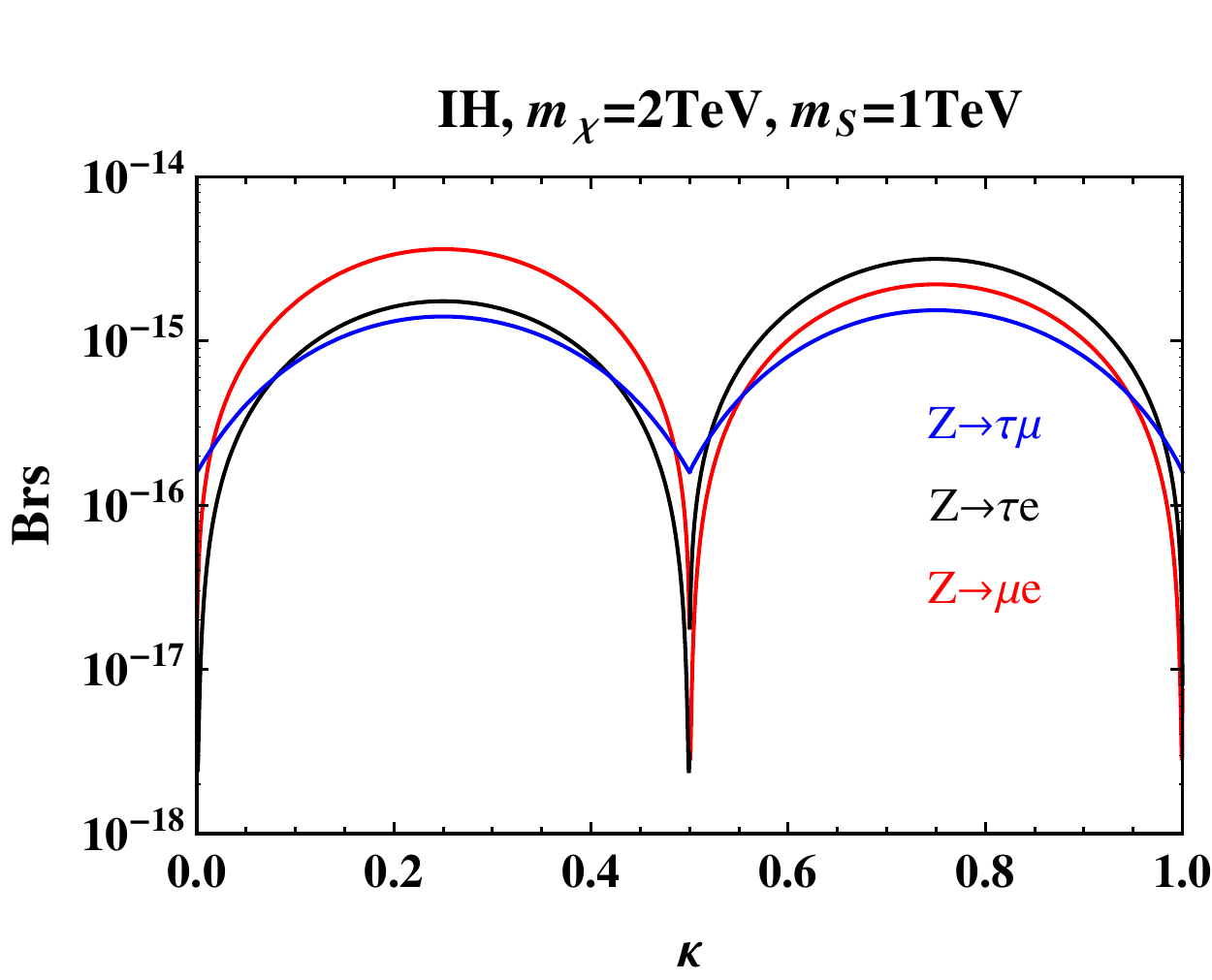}
\caption{Similar to Fig. \ref{fig5} but for the $Z$ boson decays.}
\label{fig6}
\end{figure}

By the aid of Eq. \eqref{eq:z_branching} one can numerically study the $Z$ boson LFV decays in a similar fashion. In Fig. \ref{fig6} we show their branching fractions as a function of $\kappa$. One can see that they are still well below the current experimental upper bounds. Roughly speaking, in the range of $\kappa$ not close to the cancellation points, the branching fractions follow an inverted order for NH and IH of neutrino masses.

In summary, the LFV decays of the Higgs and $Z$ bosons are severely suppressed in the color octet model especially by heavy masses of octet particles and small Yukawa couplings between them and SM leptons. They seem not to be detectable in the foreseeable near future. We will now turn to low energy LFV transitions in the next subsection.

%%%%%%%%%%%%%%
\subsection{$\mu N\to eN$}
%%%%%%%%%%%%%%

In this subsection, we will be mainly interested in the $\mu-e$ conversion in nuclei, but for the sake of comparison we will also consider the decays $\mu\to e\gamma,~3e$ by employing the analytic results in Ref. \cite{Liao:2009fm}. Since the $\mu-e$ conversion in the nucleus Ti has the best expected future sensitivity, it will be used to illustrate most of our numerical results.

The branching fractions for $\mu\to e\gamma,~3e$ are found to be \cite{Liao:2009fm},
\begin{eqnarray}
\label{eq:muegamma}
{\Br}(\mu\to e\gamma)&=&\frac{12\alpha}{\pi G_F^2m_S^4}|B_{e\mu}|^2,
\\
{\Br}(\mu\to 3e)&=&
\frac{\alpha^2}{2\pi^2G_F^2m_S^4}\bigg\{2|B+A_{e\mu}|^2+|A_{e\mu}|^2-8\textrm{Re}(BB_{e\mu}^*)
%\\
%&&
-12\textrm{Re}(A_{e\mu}B_{e\mu}^*)
+\bigg(8\ln\frac{m_\mu^2}{m_e^2}-\frac{8}{3}\bigg)|B_{e\mu}|^2\bigg\},
\label{eq:mu3e}
\end{eqnarray}
where the form factor $B$ arises from the box diagrams,
\begin{equation}\label{}
B=-\frac{\xi^2}{2\pi\alpha}z_{ex}z_{\mu x}^*z_{ey}z_{\mu y}^*H(r_x,r_y).
\end{equation}
Equation \eqref{eq:simplified_branching} implies the proportionality relation for the $\mu-e$ conversion:
\begin{equation}\label{eq:cancel}
{\Br}(\mu N\to eN)\propto|z_{ex}z_{\mu x}^*|^2[\xi F(r)+2(1-\xi)G(r)]^2m_S^{-4},
\end{equation}
where
\begin{eqnarray}
\nonumber
F(r)&=& DF_2(r)-16\sqrt{\pi\alpha}V^pF_1(r),
\\
G(r)&=& DG_2(r)-16\sqrt{\pi\alpha}V^pG_1(r).
\end{eqnarray}
The factor $|z_{ex}z_{\mu x}^*|^2$, simply summed over $x$ in the case of degenerate octet fermions, appears in all above branching fractions. It scales sensitively with the quartic coupling $\lambda_3$ between the octet scalar and the SM Higgs through Eq. \eqref{eq:para}. We will assume $\lambda_3=10^{-8}$ as in Ref.~\cite{FileviezPerez:2010ch}.

\begin{figure}[!htbp]
\centering
\includegraphics[scale=0.7]{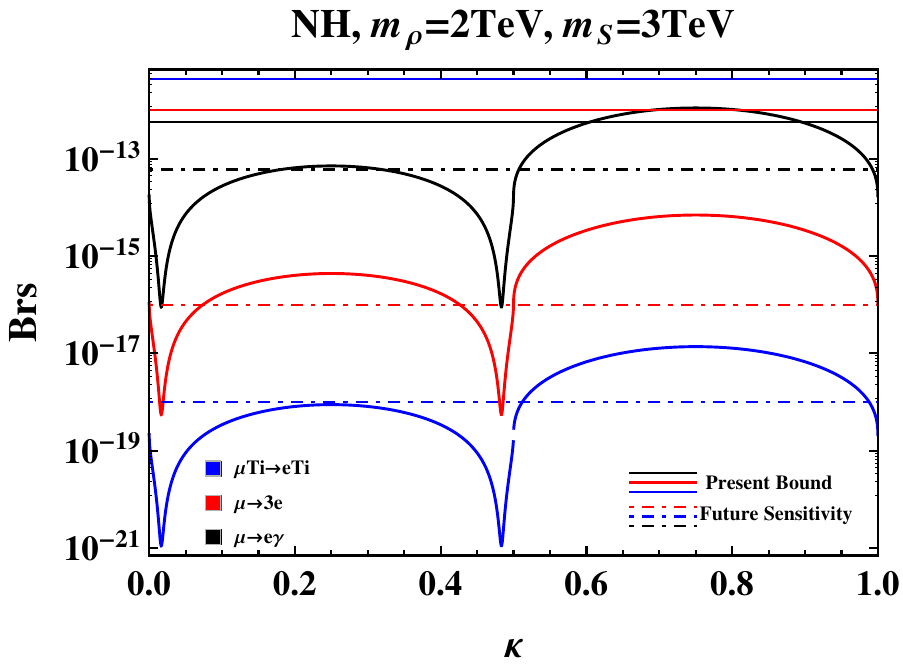}
\quad
\includegraphics[scale=0.7]{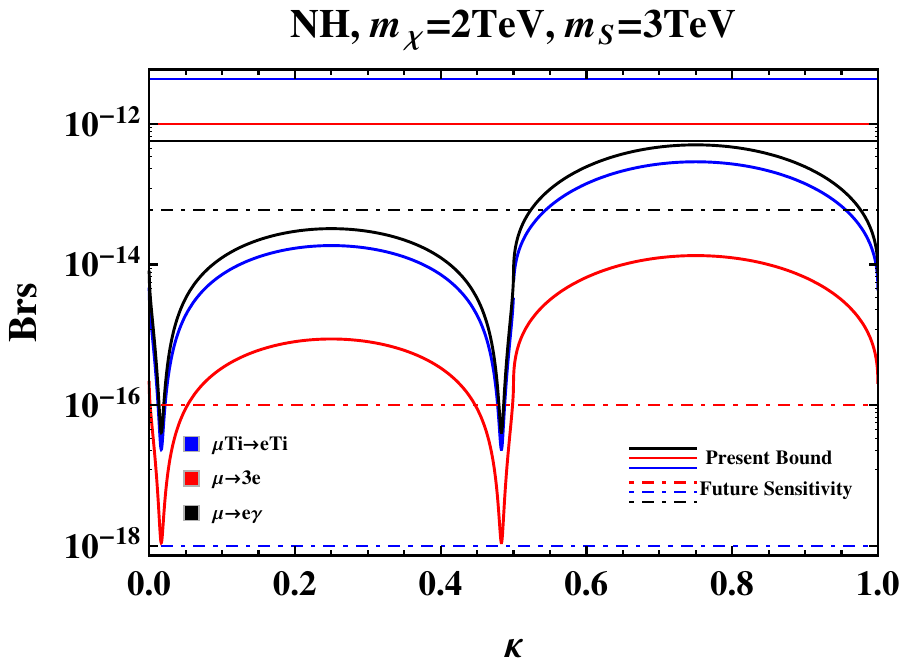}\\
\vspace{8pt}
\includegraphics[scale=0.7]{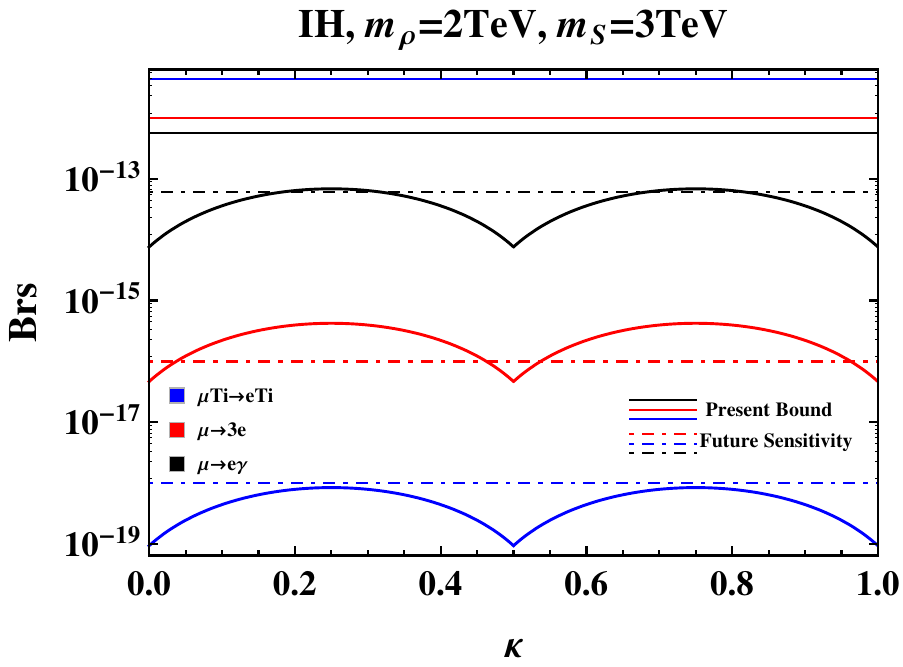}
\quad
\includegraphics[scale=0.7]{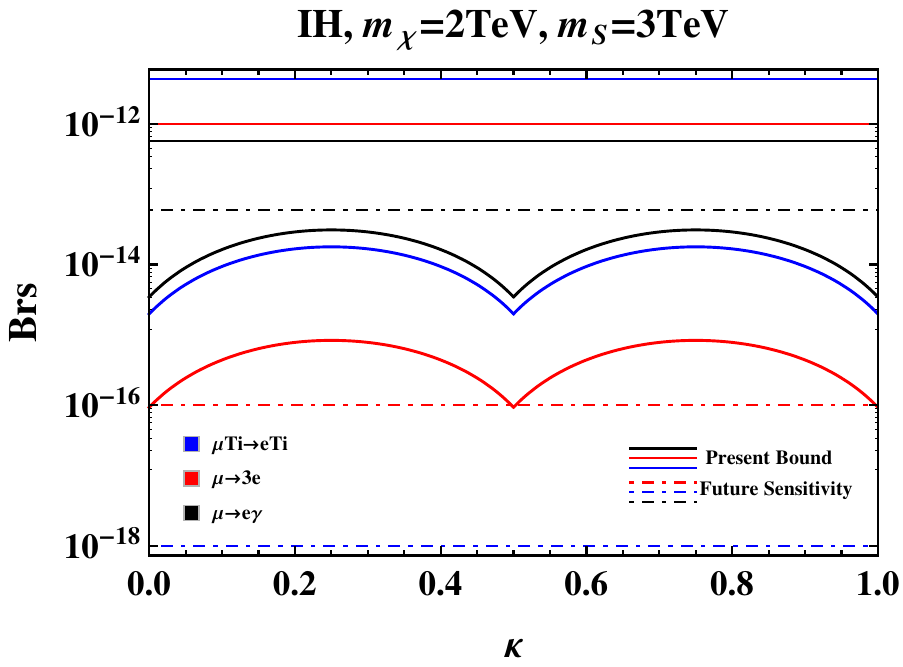}
\caption[Relation to mixing parameter]{Branching fractions for $\mu\to e\gamma,~3e$ and $\mu-e$ conversion in Ti are shown as a function of $\kappa$ in case A (left panel) and B (right) and for NH (upper panel) and IH (lower). The horizontal solid lines indicate present experimental bounds, while the dot-dashed ones are future sensitivities.}
\label{fig7}
\end{figure}

In Fig. \ref{fig7} we show the three branching fractions as a function of the $\kappa$ parameter at the fixed masses $m_{\rho(\chi)}=2~\TeV,~m_S=3~\TeV$ in case A (B) and for both NH and IH of neutrino masses. As one can see, they share the same shape and reach their extreme points at the same values of $\kappa$. This feature can be traced back to the appearance of the identical factor $|z_{ex}z_{\mu x}^*|^2$ mentioned above. We also notice that the branching fraction for $\mu-e$ conversion in the nucleus Ti in case A is about four orders of magnitude smaller than in case B. This difference arises from different combinations of form factors in Eq. \eqref{eq:cancel} for two cases.

\begin{figure}[!htbp]
\centering
\includegraphics[scale=0.7]{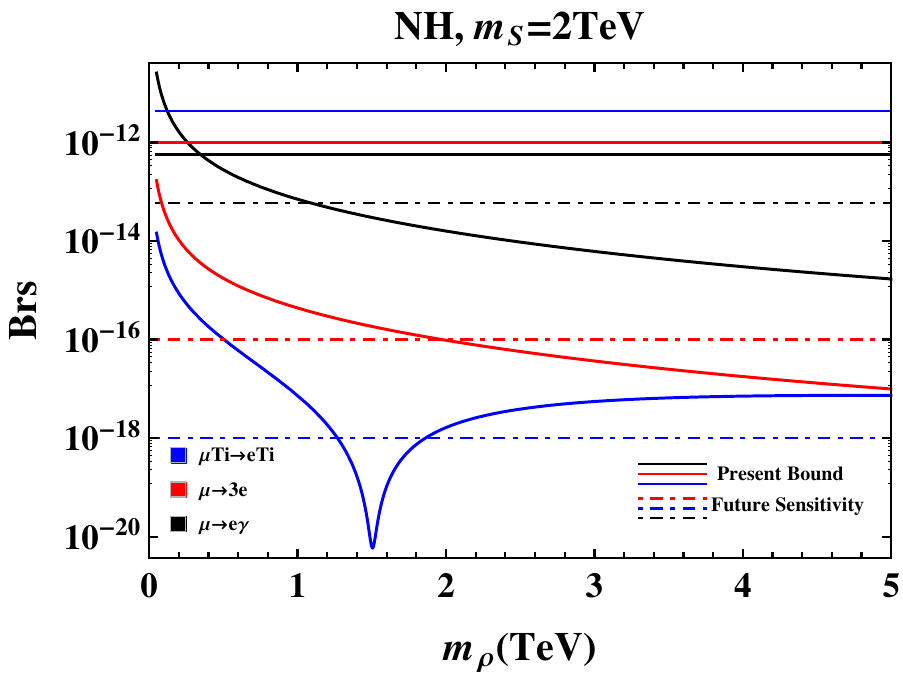}
\quad
\includegraphics[scale=0.7]{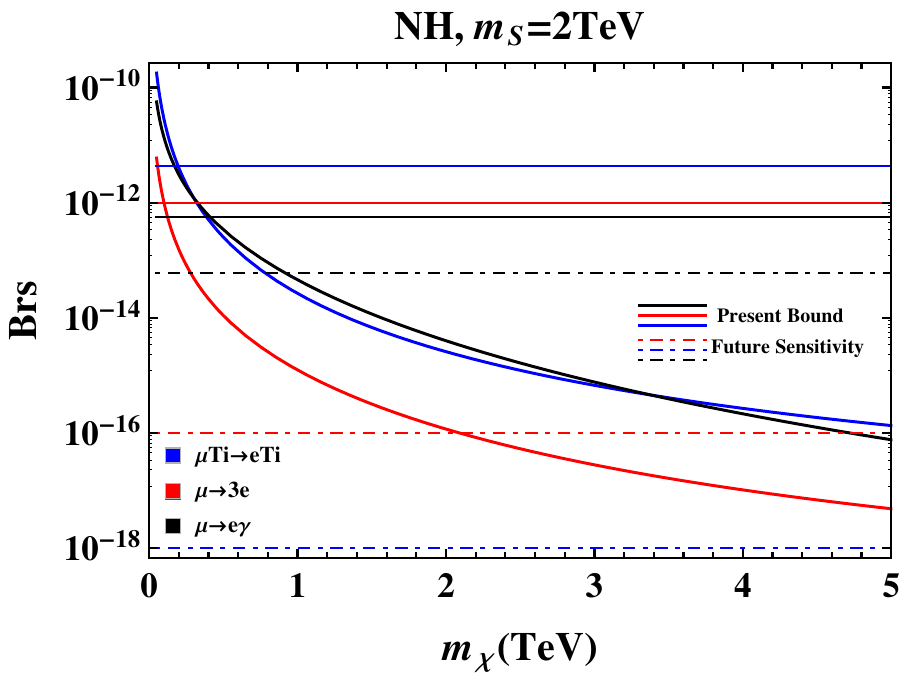}\\
\vspace{8pt}
\includegraphics[scale=0.7]{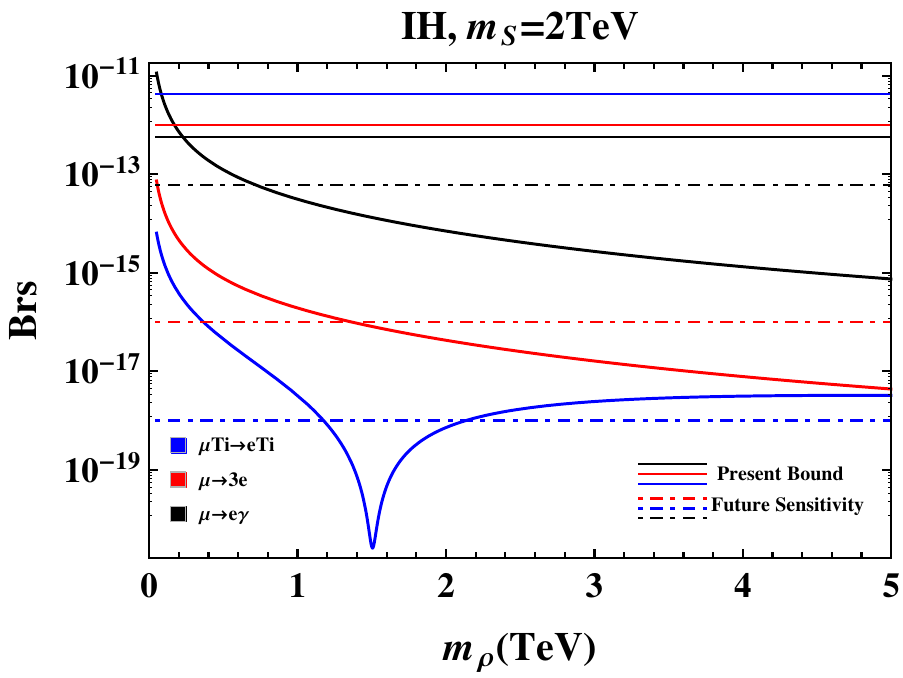}
\quad
\includegraphics[scale=0.7]{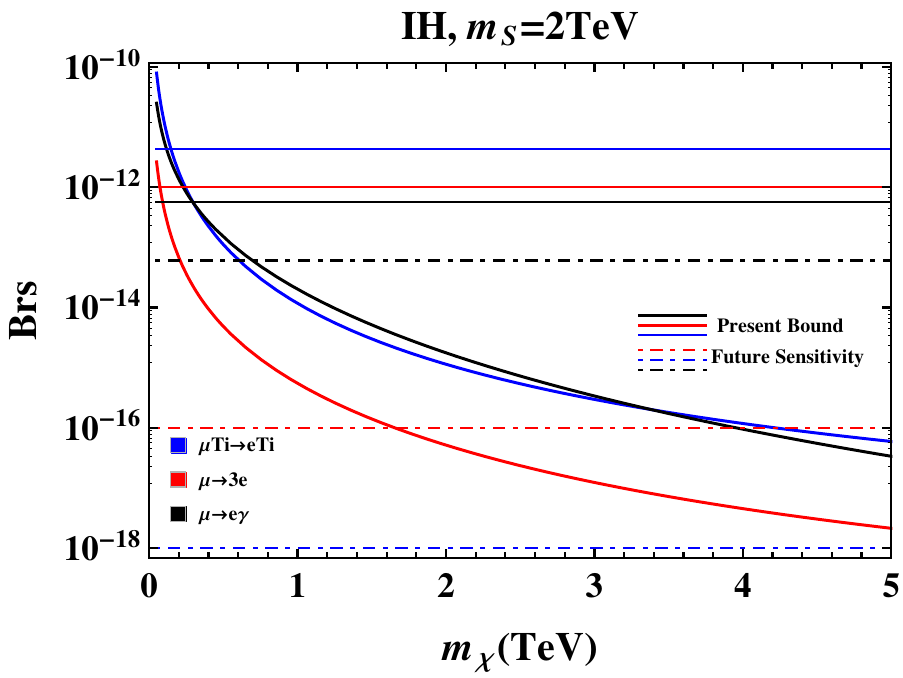}\\
\caption[Relation to mass]{Similar to Fig. \ref{fig7} but as a function of $m_{\rho(\chi)}$ for real $\omega$.}
\label{fig8}
\end{figure}

\begin{figure}[!htbp]
\setlength{\abovecaptionskip}{2pt}
\setlength{\belowcaptionskip}{2pt}
\centering
\includegraphics[scale=0.55]{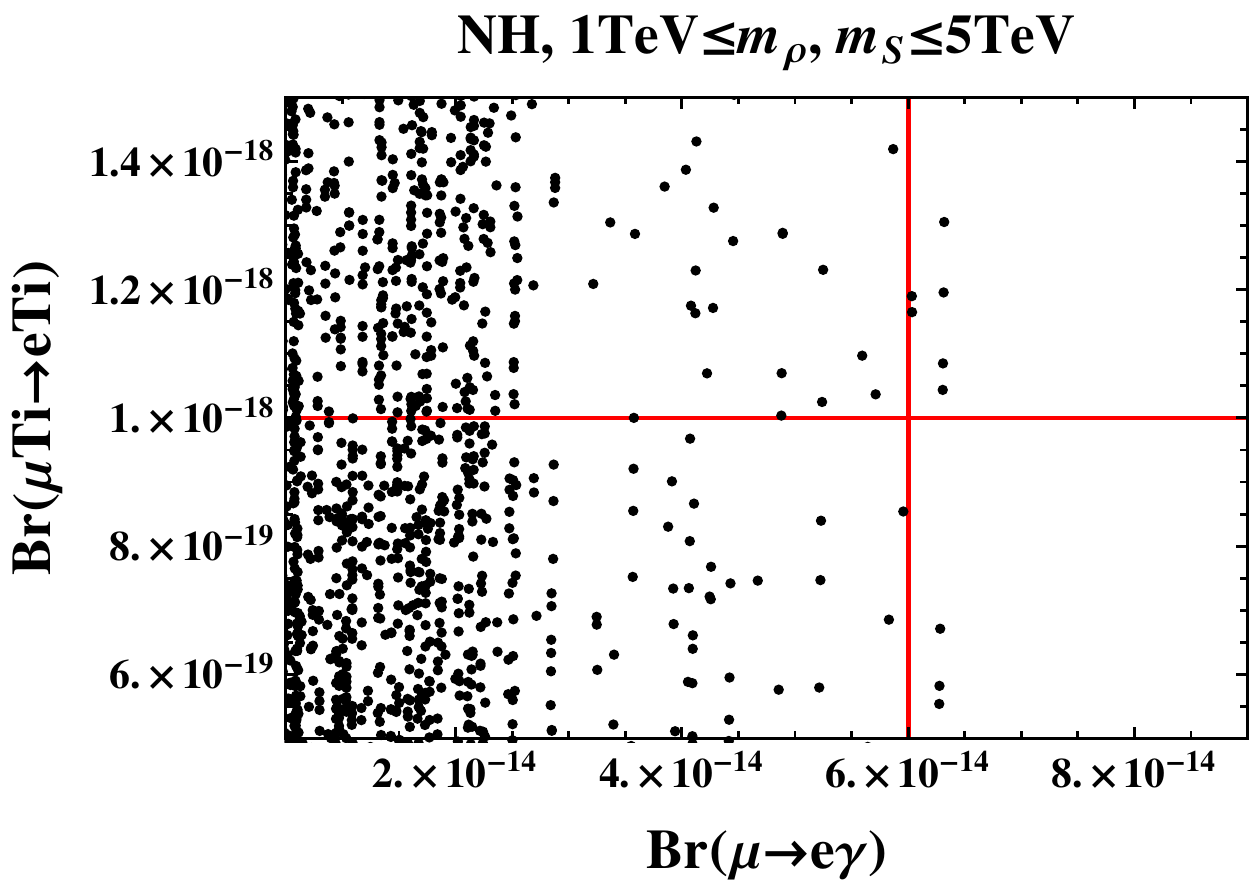}
\quad
\includegraphics[scale=0.55]{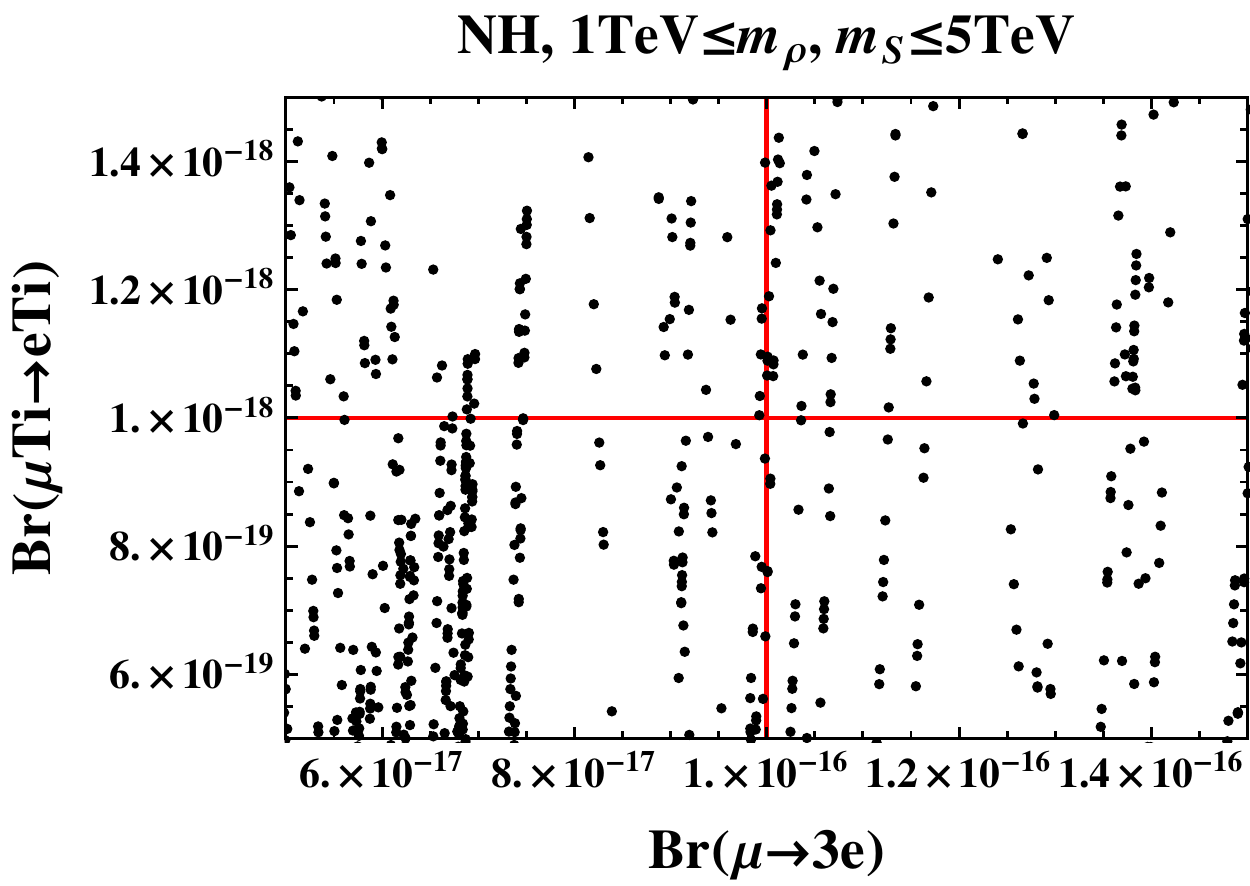}\\
\vspace{8pt}
\includegraphics[scale=0.55]{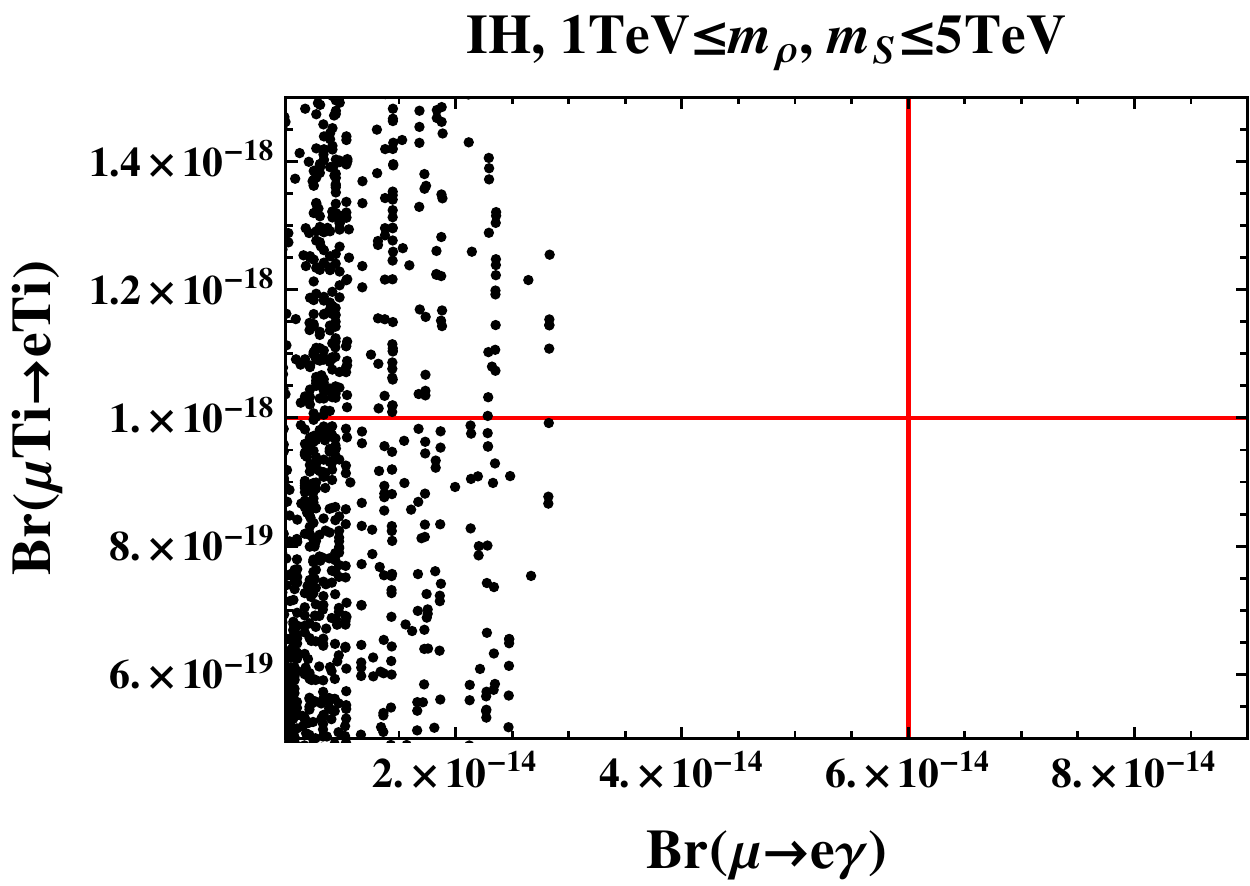}
\quad
\includegraphics[scale=0.55]{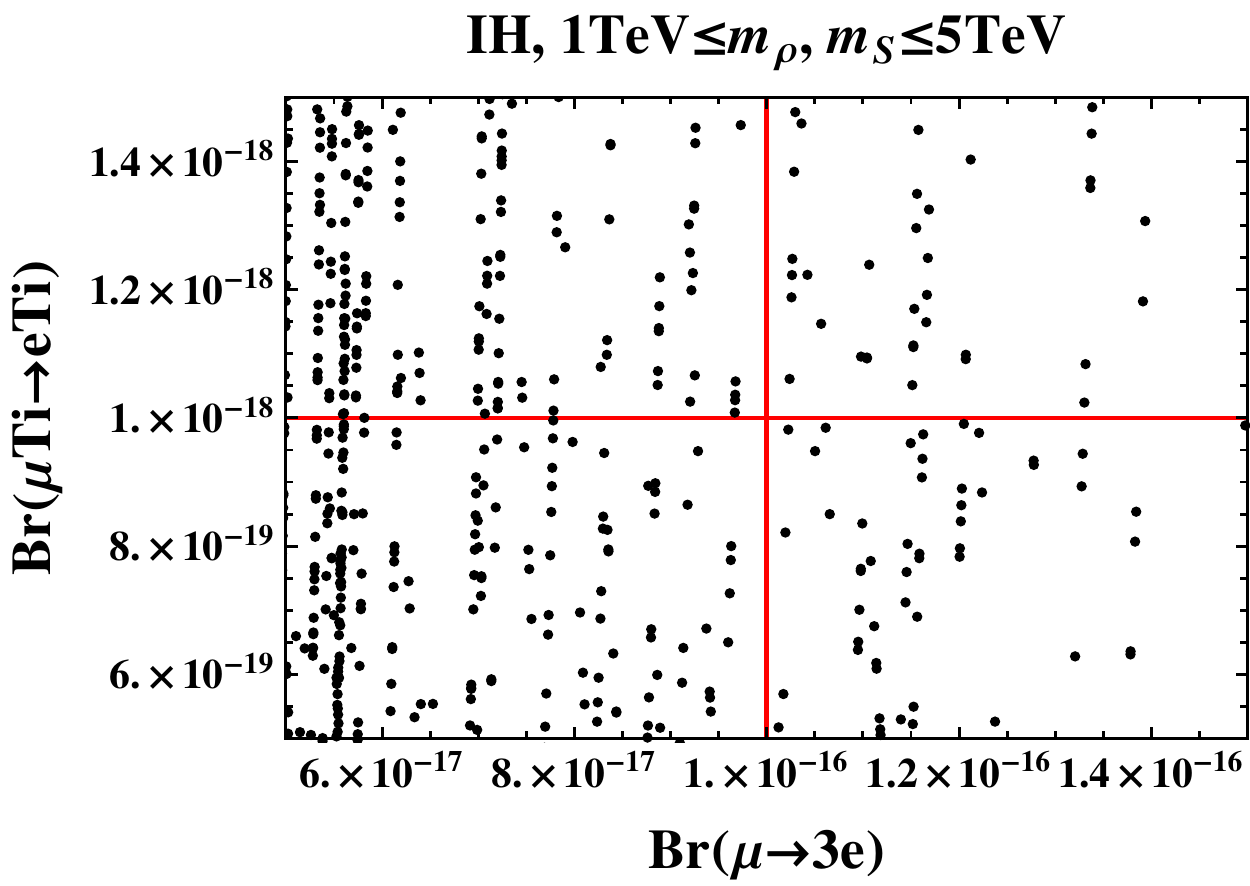}\\
\caption[Relation to mixing parameter]{Branching fractions sampled over $(m_\rho,m_S)$ in case A for NH (upper panel) and IH (lower). Solid lines indicate future sensitivities.}
\label{fig9}
\end{figure}

\begin{figure}[!htbp]
\centering
\includegraphics[scale=0.8]{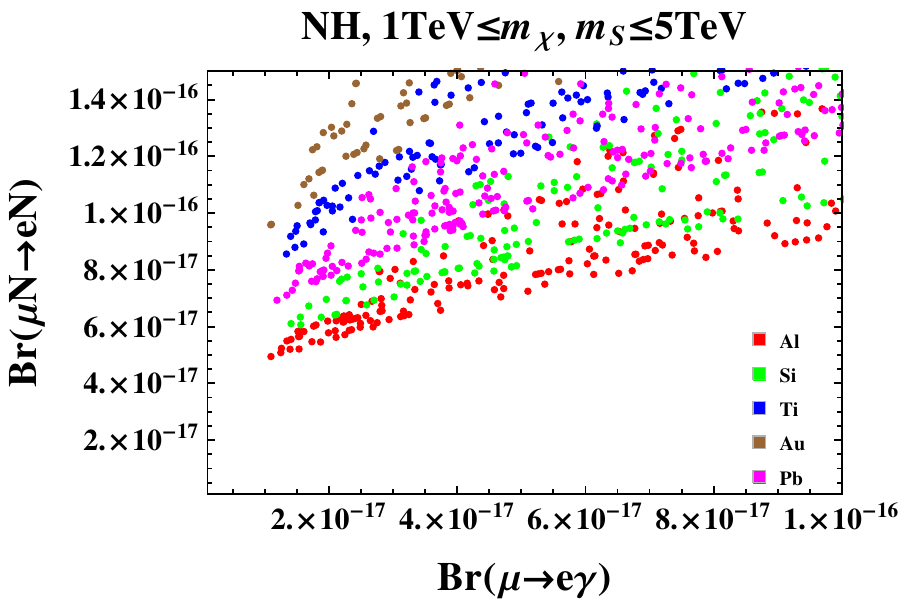}
\quad
\includegraphics[scale=0.8]{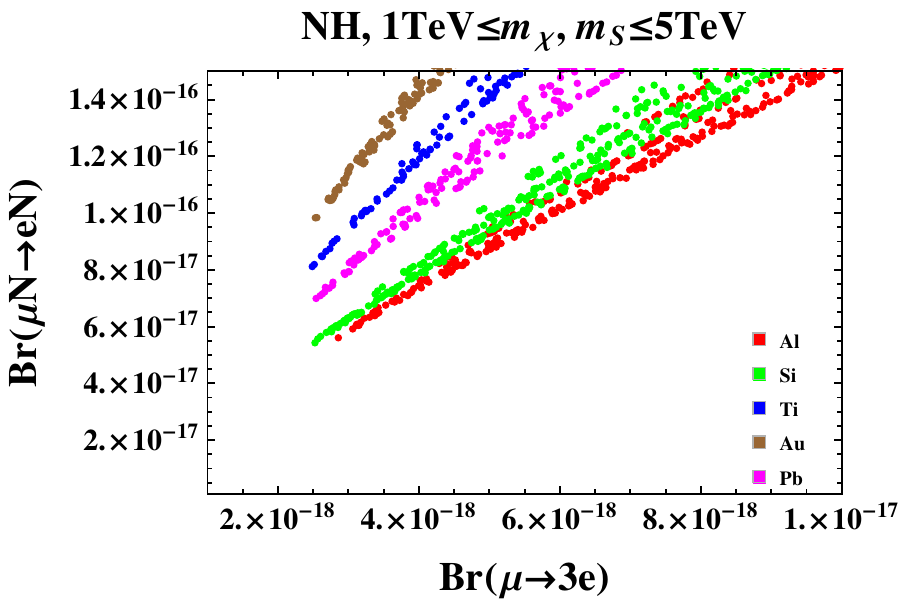}\\
\vspace{8pt}
\includegraphics[scale=0.8]{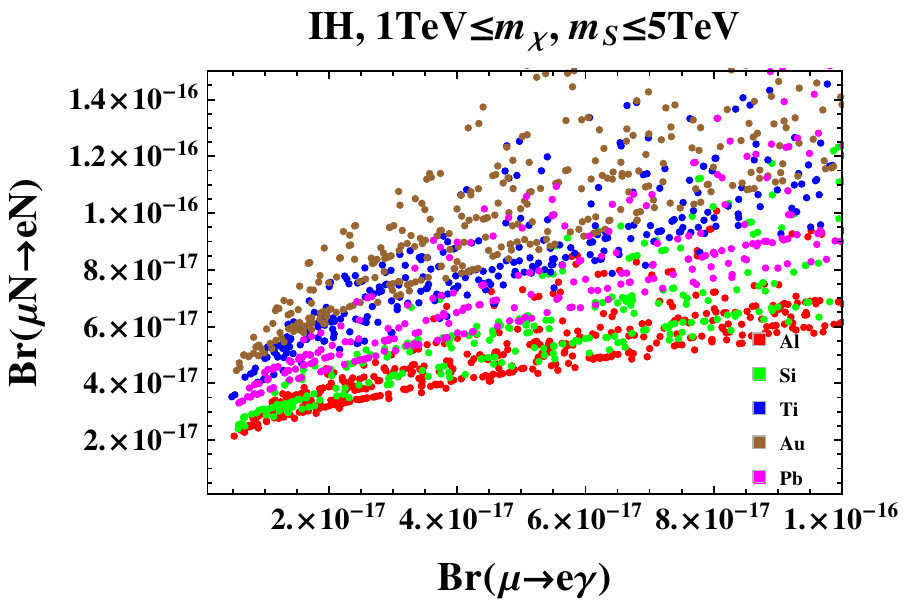}
\quad
\includegraphics[scale=0.8]{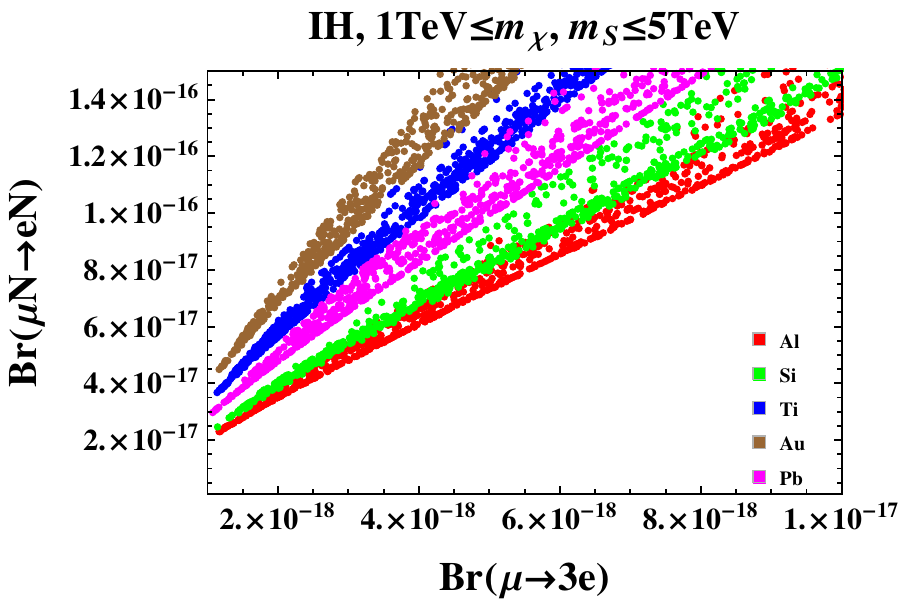}\\
\caption[Relation to mixing parameter]{Branching fractions sampled over $(m_\chi,m_S)$ in case B for NH (upper panel) and IH (lower).}
\label{fig10}
\end{figure}

Now we consider the case of a real $\omega$ parameter, in which our branching fractions become independent of it for degenerate octet fermions. Figure~\ref{fig8} shows the branching fractions as a function of $m_{\rho(\chi)}$ at fixed $m_S=3~\TeV$. We see clearly a deep dip in the branching fraction for $\mu-e$ conversion at $m_\rho\approx 1.5~\TeV$ in case A but not in case B. This can be understood by a closer look into Eq. \eqref{eq:cancel}: in case A (with $\xi=1$) the form factor $F(r)$ can vanish at a positive $r$ while in case B ($\xi=1/2$) there is no such a solution to $F(r)+G(r)=0$ for various values of $D,~V^p$. As a result, the $\mu-e$ conversion in case A can be so tiny in some regions of parameter space that it could even evade future sensitivities. To show this more explicitly, in Fig. \ref{fig9} we scan over a larger set of parameters by sampling over $m_S,~m_\rho$ from $1~\TeV$ to $5~\TeV$ in case A. In fact, the dip of $\mu-e$ conversion arises essentially from the cancellation between the anapole ($F_1$) and dipole ($F_2$) terms of $\mu\to e\gamma^*$ as they contribute oppositely to the form factor $F(r)$. Since its branching fraction can easily meet the current bounds, future experiments will be important to constrain the range of masses. To assess whether case B can also evade future sensitivities, we do the same sampling in Fig. \ref{fig10}. As is illustrated, the Yukawa couplings in this scenario are essentially determined by the low energy neutrino parameters, which leads to fairly strong correlations among these processes and in particular between $\mu\to 3e$ and $\mu-e$ conversion in nuclei. Since the future sensitivity of $\mu-e$ conversion in Ti is expected to reach a level of $10^{-18}$, it will be capable of excluding case B in the scanned regions of parameter space while  reserving significant portions of parameter space in case A. It is worth recalling that case B still survives the present constraints.

%\vspace{1cm}

\begin{figure}[!htbp]
\centering
\includegraphics[scale=0.7]{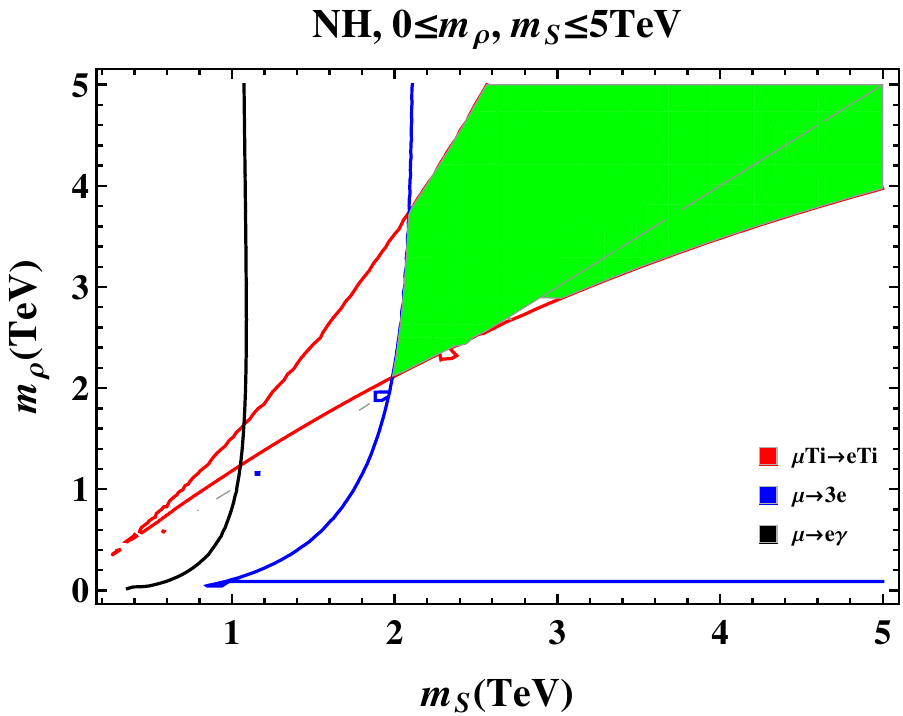}
\quad
\includegraphics[scale=0.7]{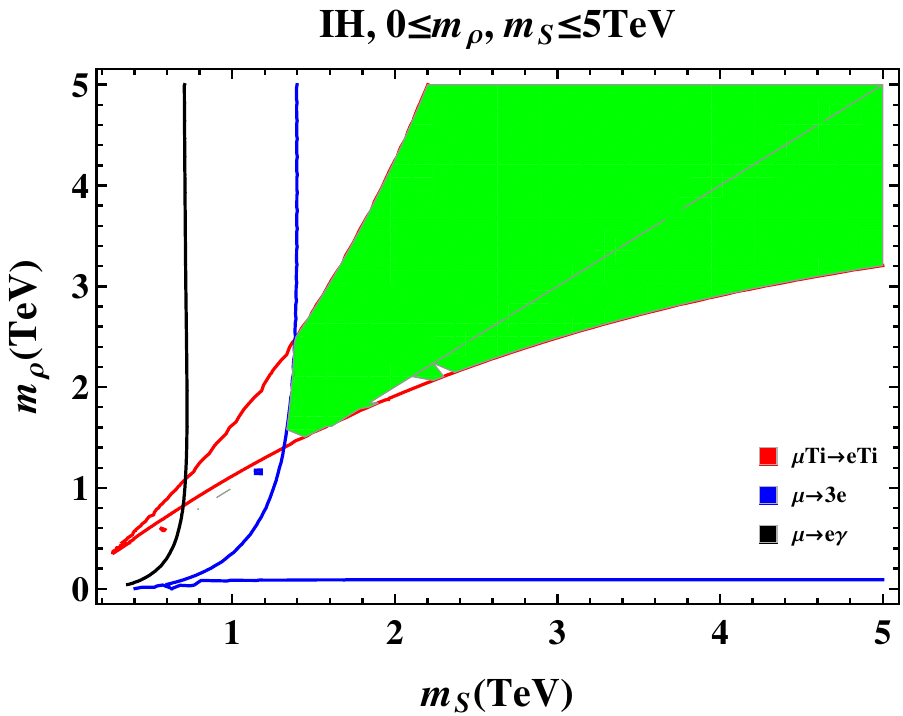}\\
\caption[Relation to mass]{Contours of $\Br(\mu\text{Ti}\to e\text{Ti})$, $\Br(\mu\to 3e)$, and $\Br(\mu\to e\gamma)$ in $m_S-m_\rho$ plane assuming future sensitivities.}
\label{fig11}
\end{figure}

Finally we show in Fig. \ref{fig11} the contours of $\Br(\mu\text{Ti}\to e\text{Ti})$, $\Br(\mu\to 3e)$, and $\Br(\mu\to e\gamma)$ in the $m_\rho-m_S$ plane, for a real $\omega$ in case A and using the best-fit values of neutrino oscillation parameters. The red, blue, and black curves denote future experimental sensitivities, and the green region denotes parameter space not to be excluded by these limits. In the long term the decay $\mu\to 3e$ and $\mu-e$ conversion in nuclei will be more stringent than $\mu\to e\gamma$. These experiments are expected to set relevant constraints on $m_\rho,~m_S$ and rule out relatively low-mass regions. A rough estimate of the lower bounds turns out to be:
\begin{eqnarray}\label{eq:mass_lower}
\nonumber
\text{NH} &&m_S>2.0\text{TeV}, \;\;m_\rho>2.0\text{TeV},
\\
\text{IH} &&m_S>1.4\text{TeV}, \;\;m_\rho>1.5\text{TeV}.
\end{eqnarray}

%%%%%%%%%
\section{Summary}
%%%%%%%%%
In this work we have investigated systematically the LFV phenomenology of the color octet model, covering the LFV decays of the Higgs and $Z$ bosons and the $\mu-e$ conversion in nuclei. For the latter we have taken into account both photonic and non-photonic contributions and found that the latter is indeed subdominant. As the flavor structure in the Yukawa couplings between the SM leptons and the color octet particles is mainly determined by neutrino oscillation data, the couplings can be expressed in terms of very few free parameters which could be constrained by various LFV observables. Currently, the LFV bounds on the model are not stringent enough; however, future experiments with impressive expected sensitivity will be capable of probing larger portions of the parameter space and strongly constraining the masses of octet particles. As a consequence of cancellation between the anapole and dipole terms in the photonic contribution, a large portion of parameter space in case A can even survive the future sensitivity for $\Br(\mu N\to eN)$. On the other hand, the triplet case of fermions (case B) is expected to be excluded by future limit of $\Br(\mu \text{Ti}\to e\text{Ti})<10^{-18}$. In the foreseeable future, these low energy LFV transitions can give better constraints than the Higgs and $Z$ boson decays at high energy colliders, and can serve as a valuable addition to direct collider searches for new particles.

%\newpage
\vspace{0.5cm}
\noindent %
\section*{Acknowledgement}

This work was supported in part by the Grants No. NSFC-11025525, No. NSFC-11575089 and by the CAS Center for Excellence in Particle Physics (CCEPP).

\appendix
\section{Some values used}\label{app:A}
The coefficients $G_S$s for scalar operators in Eq. \eqref{eq:coefficients} are introduced in Ref.~\cite{Kitano:2002mt}:
\begin{eqnarray}%\label{}
\nonumber
&&G_S^{p,u}=G_S^{n,d}=5.1,\quad G_S^{p,d}=G_S^{n,u}=4.3,\quad G_S^{p,s}=G_S^{n,s}=2.5,
\\
&&G_V^{p,u}=G_V^{n,d}=2.0,\quad G_V^{p,d}=G_V^{n,u}=1.0,\quad G_V^{p,s}=G_V^{n,s}=0.
\end{eqnarray}
The overlap integrals $D,~S^{p},~S^{n},~V^{p},~V^{n}$ are related to nuclear physics, and recorded here in Table \ref{tab:overlap} together with $\mu$ capture rates $\Gamma_\textrm{capt}$ for various nuclei \cite{Kitano:2002mt}.
\begin{table}[!hbp]
\centering
\begin{tabular}{|c|c|c|c|c|c|c|}
%\hline
\hline
Nucleus& $D$ & $S^{p}$ & $S^{n}$ & $V^{p}$&$V^{n}$&$\Gamma_\textrm{capt}(10^6s^{-1})$\\
\hline
$\quad_{\text{13}}^{\text{27}}\text{Al}$ & 0.0362 & 0.0155 & 0.0167 & 0.0161&0.0173&0.7054\\
$\quad_{\text{14}}^{\text{28}}\text{Si}$&0.0419&0.0179&0.0179&0.0187&
0.0187&0.8712\\
$\quad_{\text{22}}^{\text{48}}\text{Ti}$ & 0.0864 & 0.0368 & 0.0435 & 0.0396&0.0468&2.59\\
$\quad_{\text{79}}^{\text{197}}\text{Au}$ & 0.189 & 0.0614 & 0.0918 & 0.0974&0.146&13.07\\
$\quad_{\text{82}}^{\text{208}}\text{Pb}$ & 0.161 & 0.0488 & 0.0749 & 0.0834&0.128&13.45\\
\hline
\end{tabular}
\caption{Nuclear form factors in units of $m_\mu^{5/2}$ and capture rates for various nuclei.}\label{tab:overlap}
\end{table}

\section{Loop functions}\label{app:B}
The loop functions used in our calculation are:
\begin{eqnarray}
\nonumber
F(x)&=&\frac{1}{4(1-x)^3}(1-4x+3x^2-2x^2\ln{x}),~F(1)=\frac{1}{6},
\\
\nonumber
F_1(x)&=&\frac{1}{36(1-x)^4}(2-9x+18x^2-11x^3+6x^3\ln{x}),~F_1(1)=\frac{1}{24},
\\
\nonumber
F_2(x)&=&\frac{1}{12(1-x)^4}(1-6x+3x^2+2x^3-6x^2\ln{x}),~F_2(1)=\frac{1}{24},
\\
\nonumber
G_1(x)&=&\frac{1}{36(1-x)^4}(-16+45x-36x^2+7x^3-12\ln{x}+18x\ln{x}),~G_2(1)=\frac{1}{8},
\\
\nonumber
G_2(x)&=&\frac{1}{12(1-x)^4}(-2-3x+6x^2-x^3-6x\ln{x}),~G_2(1)=-\frac{1}{24},
\\
\nonumber
H(x,y)&=&\frac{1}{4(x-y)}\bigg[\frac{1}{1-x}-\frac{x\ln{x}}{1-x}+\frac{x\ln{x}}{(1-x)^2}
-\frac{1}{1-y}+\frac{y\ln{y}}{1-y}-\frac{y\ln{y}}{(1-y)^2}\bigg],
\\
\nonumber
H(x,x)&=&\frac{1}{4(1-x)^3}(1-x^2+2x\ln x),~H(1,1)=\frac{1}{12},
\\
H(x,0)&=&\frac{1}{4(x-1)^2}(1-x+x\ln x)\equiv H(x), ~H(1)=\frac{1}{8}.
\end{eqnarray}

\end{document}